\documentclass[aps,twocolumn,showpacs,superscriptaddress,amsfont,graphicx,nofootinbib,preprintnumbers]{revtex4-2}%
\UseRawInputEncoding

\usepackage{amsmath}
\usepackage{cases}
\usepackage{amssymb}
\usepackage{amsfonts}
\usepackage{dcolumn}
\usepackage{bm}
\usepackage{epsfig}
\usepackage{bbm}
\usepackage{graphicx}
\usepackage{xcolor}
\usepackage{array}
\usepackage{subfigure}
\usepackage{hyperref}
\usepackage{mathrsfs}
\usepackage{verbatim}
\usepackage{float}
\usepackage{epstopdf}
\usepackage{color}

\newcommand{\gsim}{\mathrel{\hbox{\rlap{\lower.55ex \hbox {$\sim$}}
			\kern-.3em \raise.4ex \hbox{$>$}}}}
\newcommand{\lsim}{\mathrel{\hbox{\rlap{\lower.55ex \hbox {$\sim$}}
			\kern-.3em \raise.4ex \hbox{$<$}}}}

\hypersetup{colorlinks=true,
	breaklinks=true,
	pdfstartview=Fit,
	linkcolor=blue,
	citecolor=blue,
	urlcolor=blue}

\begin{document}
\title{Probing levitodynamics with multi-stochastic forces 
and the simple applications on the dark matter detection in 
optical levitation experiment}
\author{Xi Cheng}
\affiliation{Beijing University of Technology, Beijing 100124, China}

\author{Ji-Heng Guo}
\email{Guojiheng@buaa.edu.cn}
\affiliation{School of Physics, Beihang University, Beijing 100083, China}

\author{Wenyu Wang}
\email{wywang@bjut.edu.cn}
\affiliation{Beijing University of Technology, Beijing 100124, China}

\author{Bin Zhu}
\email{zhubin@mail.nankai.edu.cn}
\affiliation{School of Physics, Yantai University, Yantai 264005, China}
\begin{abstract}
If the terrestrial environment is permeated by dark matter, the levitation experiences damping forces and fluctuations attributed to dark matter. This paper investigates levitodynamics with multiple stochastic forces, including thermal drag, photon recoil, feedback, etc., assuming that all of these forces adhere to the fluctuation-dissipation theorem. The ratio of total damping to the stochastic damping coefficient distinguishes the levitodynamics from cases involving only one single stochastic force. The heating and cooling processes are formulated to determine the limits of temperature change. All sources of stochastic forces are comprehensively examined, revealing that dark matter collisions cannot be treated analogously to fluid dynamics. Additionally, a meticulous analysis is presented, elucidating the intricate relationship between the fundamental transfer cross-section and the macroscopic transfer cross-section. While the dark damping coefficient is suppressed by the mass of the levitated particle, scattering can be coherently enhanced based on the amount of the component microscopic particle, the nucleus form factor, and the static structure factor. Hence, dark damping holds the potential to provide valuable insights into the detection of the macroscopic strength of fundamental particles. We propose experimental procedures for levitation and employ linear estimation to extract the dark damping coefficient. Utilizing current levitation results, we demonstrate that the fundamental transfer cross section of dark matter can be of the order $\sigma^{\rm D}_{T}\lsim {\cal O}(10^{-26})\rm cm^2$.
\end{abstract}


\maketitle

\section{Introduction}

Observations in cosmology and astrophysics have confirmed the presence of dark matter (DM)~\cite{Bertone:2004pz}. One of the most promising experimental approaches to detect DM on Earth is to search for small energy depositions resulting from elastic scattering of DM in sensitive detectors. Searches for weakly interacting massive particles (WIMPs) represent some of the most advanced techniques in this field. However, stringent constraints exist on the cross section for DM particles with masses exceeding 1 GeV. DM particles may have eluded detection if the resulting energy deposits fall below the threshold of current detectors. Therefore, it is imperative to explore alternative strategies and develop novel detection methods to investigate the remaining parameter space of DM i.e., light DM~\cite{migdal1939ionizatsiya,Vergados:2005dpd,Moustakidis:2005gx,Ejiri:2005aj,Bernabei:2007jz,Ibe:2017yqa,Dolan:2017xbu,PhysRevD.101.015012,Baxter:2019pnz,Essig:2019xkx,GrillidiCortona:2020owp,Knapen:2020aky,Liang:2020ryg,Flambaum:2020xxo,Bell:2021zkr,Acevedo:2021kly,Wang:2021oha,Guo:2021imc,Riedel:2012ur}.

In recent years, a fundamentally different approach has been proposed to increase the sensitivity to the energy transfer from DM particles to specific microscopic internal degrees of freedom within a detector. This approach involves optically monitoring the center-of-mass (COM) motion of a levitated macroscopic object. The levitation of micrometer-sized objects in vacuum was first demonstrated in the pioneering work of Arthur Ashkin in the 1970s~\cite{PhysRevLett.24.156}. 
Levitated nanoparticles have since been utilized in various fields, 
including trapping and cooling of atoms, control mechanisms 
based on optical, electrical, and magnetic forces, 
investigations of light-matter interactions,
and detection of dark forces
~\cite{Gieseler_2014,PhysRevA.91.051805,
2013NatPh...9..806G,PhysRevLett.112.103603,neukirch2015multi1,kiesel2013cavity,PhysRevLett.114.123602,Chang:2009buc,
PhysRevA.83.013803,kaltenbaek2012macroscopic,Arvanitaki:2012cn,winstone2023levitated,PhysRevApplied.15.014050,Yin:2022geb}.

Laboratory-scale levitation experiments aiming to reach the quantum regime require a substantial degree of isolation and control, making them highly sensitive to forces or phenomena that couple to mass. Recently, a study by Ref.~\cite{Afek:2021vjy} demonstrated that levitated nanoscale mechanical devices operated around the standard quantum limit for impulse sensing can detect the deposition of energy transfer from DM particles. Levitation devices offer a unique opportunity for directional searches for DM masses in the keV-GeV regime, which currently have limited direct detection constraints. Moreover, they exhibit remarkable complementarity to other proposals due to coherent interaction at low momentum detection thresholds~\cite{Monteiro:2020wcb}. The results of Ref.~\cite{Monteiro:2020wcb} demonstrate the potential of optomechanical sensors in searching for DM. Large arrays of sensors could facilitate the reconstruction of track-like signals from DM particles while possessing sufficient sensitivity to detect and count individual collisions of latent gas in ultra-high vacuum environments. Cavity optomechanical systems aim to operate in the shot-noise dominant regime to observe macroscopic quantum phenomena.

In this study, we focus on the detection of DM using levitated particles and investigate the thermal dynamics of Langevin systems subjected to multi-stochastic forces. The presence of distinct stochastic forces can induce fluctuation-dissipation from an unknown source, making it challenging to detect the DM-induced collision signal amidst the noise. By examining the properties of these stochastic forces and studying the recoil of levitated particles, we aim to unveil their characteristics and provide insights into the presence of additional unique sources. The rest of this paper is organized as follows: Sec. \ref{sec2-th} presents the general formulation of the Langevin system with multi-stochastic forces. Sec. \ref{sec3-st} examines the properties of various stochastic forces. Sec. \ref{sec4-dm} discusses the application of these forces in detecting dark matter. Finally, Sec. \ref{sec5-cl} concludes the paper.

\section{Levitodynamics with multi-stochastic forces}
\label{sec2-th}

\subsection{Why we go beyond the framework of fluctuation theorem}

\begin{figure*}[htpb]
\begin{center}
\includegraphics[width=140mm]{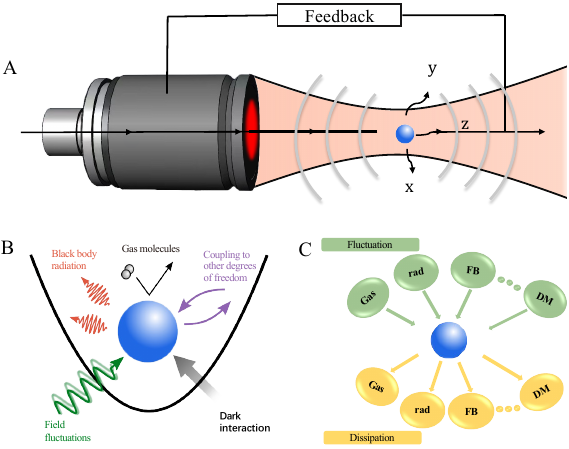}
\caption{ A: Sketch map of the levitation  of particle 
conducted by laser cooling techniques, The feedback loop
can be turned on or off in the experiments. B: The
noise sources in levitodynamics
\footnote{The figure is adapted from Fig. 4 of Ref.~\cite{Gonzalez-Ballestero:2021gnu}}: (i) collisions with molecules of the 
surrounding gas; (ii) emission, absorption, and scattering of thermal 
electromagnetic radiation; (iii) noise in the trap center from 
vibrations and/or stray fields;  (iv) noise in the trap frequency from
fluctuations of the electromagnetic fields that generate the trapping 
potentials;  (v) cross-coupling to other thermal 
degrees of freedom in the particle either motion along orthogonal 
directions, rotation, or internal excitation; and (vi) the unknown
interaction which can be dark interaction talked about in this paper. 
C: The fluctuations and dissipation of the Langevin system
with multi-stochastic sources. The arrows show the energy flow.
}\label{sec2-lv1}
\end{center}
\end{figure*}

Nanoparticle levitation or trapping can be achieved through laser cooling techniques, and usually there are two techniques used to cool macroscopic mechanical systems, one is passively using optical tweezers with cavity composited of reflective mirrors ~\cite{opttweezer2020}, and the other is utilizing the parametric ``feedback" forces control as adjusting the damping force on the levitated particle according to its position as illustrated in panel A of Fig.~\ref{sec2-lv1} ~\cite{Gieseler_2014}. By enabling or disabling the feedback loop, the heating (relaxation) or cooling processes of the levitated particle can be measured. Apart from the feedback forces, the levitated particle can interact with the environment through various other forces, as illustrated in panel B of Fig.~\ref{sec2-lv1}
~\cite{Gonzalez-Ballestero:2021gnu}.
Noise in the trapping potentials generates fluctuations and dissipation in particle motion. In the quantum regime, these mechanisms induce decoherence in the motional quantum state by leaking information about the particle's position to the environment. The primary sources of noise in levitodynamics include (i) collisions with gas molecules in the vicinity; (ii) emission, absorption, and scattering of thermal electromagnetic radiation; (iii) noise in the trap center due to vibrations and/or stray fields; (iv) noise in the trap frequency caused by fluctuations in the electromagnetic fields generating the trapping potentials;  (v) cross-coupling to other thermal degrees of freedom in the particle's motion along orthogonal directions, rotation, or internal excitation; and (vi) the unknown
interaction which can be dark interaction talked about in this paper. Further analysis of these noise sources can provide a deep understanding of levitated particles and enhance sensor accuracy. In the next section, we will explore the thermal dynamics of this noise.

Although the system may have numerous degrees of freedom, a good approximation is that these freedoms are well decoupled. Each degree of freedom exhibits independent harmonic oscillator motion with frequency $\omega_0$ and damping coefficient $\gamma$. The system can be described by a one-dimensional Brownian particle with mass $m$ following the equation of motion
\begin{equation}
\frac{d^{2}x}{dt^{2}}+\gamma\frac{dx}{dt}+\omega_{0}^{2}x=\frac{F\left(t\right)}{m}\, .\label{laneq}
\end{equation}
It is known as the Langevin system. The force $F(t)$ can be a time-dependent external force used to explore response functions or a time-dependent random force resulting from molecular collisions with the environment. For example, in the case of Brownian motion in a fluid, $\gamma$ represents the damping coefficient of spherical particles. The stochastic force $F(t)$ has an expectation value $\langle F\rangle=0$. Its correlation function is given by
\begin{equation}
\langle F(t)F(t^{\prime}) \rangle = 2m \gamma k_{\rm B} T \delta (t-t^{\prime}),
\label{general-core}
\end{equation}
where $k_{\rm B}$ is the Boltzmann constant and it is proportional to the temperature of the environment $T_{\rm E}$. The system establishes a fundamental relationship between fluctuation and dissipation. The fluctuating force is related to the coefficient of viscosity of the fluid, representing dissipative forces in the system, and it depends on the temporal characteristics of molecular fluctuations.
The energy of the levitated particle changes due to its interaction with the environment, and the time evolution of the average energy $\langle E\rangle$ is described by the Fokker-Planck equation. The energy flow can be understood as fluctuations driving the levitated particle while dissipation dampens it. In the equilibrium state, the energy from both sides compensates each other.
An effective temperature, equal to the temperature of the environment, can be assigned to the system.

However, as mentioned earlier, the levitated particle interacts with multiple stochastic forces simultaneously, resulting in the presence of multiple force sources. In particular, these different stochastic forces don't need to be in equilibrium, which leads to the failure of the theoretical formulation mentioned above. Nevertheless, not every damping coefficient needs to correspond to a single fluctuation source, thereby altering the thermal dynamics of the levitated particle. While it is possible to achieve a steady state for the levitated particle, this state does not indicate the temperature of the environment due to the absence of a uniform temperature. For instance, in the levitation  system, the temperature in the correlation function of the feedback cooling force (as described by Eq.~\eqref{general-core}) must be significantly lower than the environment temperature due to collisions with surrounding gas molecules. Otherwise, the cooling process will not be effective. Consider the possible dark matter interaction studied in this paper, where the dark matter decoupled from the heat bath during the early epoch of the universe. There is no reason for the correlation function of the dark force to be identical to that of the gas environment. The details of the Langevin system with multiple stochastic forces should be investigated. In such scenarios, a steady state can be achieved, and the fluctuation and dissipation can be observed, as shown in panel C of Fig.~\ref{sec2-lv1}. The levitated particle is driven by fluctuations from various sources and dissipates through similar multi-sources. While the total energy flow in the fluctuations is equal to that of the dissipation in steady states, this relationship may not hold for individual pairs of fluctuations and dissipation.

\subsection{New framework}
In this work, we extend the framework of fluctuation theorems, as introduced in Ref.~\cite{2014NatNa...9..358G}, to investigate the Langevin system with multiple stochastic forces. This framework enables the study of relaxation from a non-equilibrium state towards equilibrium. The trapped particle experiences a trap force generated by the laser as well as stochastic forces arising from random impacts of gas molecules, radiation, and dark interactions. As demonstrated in Ref.~\cite{2014NatNa...9..358G}, the platform established in this study enables experimental investigations of non-equilibrium fluctuation theorems for arbitrary steady states. Furthermore, it can be extended to explore quantum fluctuation theorems and systems that do not adhere to detailed balance. To simplify the analysis, we focus on four specific types of forces:
\begin{enumerate}
\item The force arises from collisions between the levitated particle and the thermally moving molecules in the atmosphere. This force is characterized by a damping coefficient $\gamma_{\rm TH}$ and a stochastic driving force $F_{\rm TH}$, as shown in Eq.~\eqref{general-core}. The damping coefficient can be obtained from the collisions of microscopic components in fluid dynamics.
\item The heating rate is attributed to the stochastic driving force $F_{\rm RE}$ arising from photon recoil kicks. Additionally, a damping coefficient $\gamma_{\rm RE}$ (denoted as $\gamma_{\rm \epsilon }$ in Ref.~\cite{Jain2016DirectMO}) arises from the back-reaction of the scattered field on the motion of the levitated particle and can be evaluated accordingly.~\cite{Millen:2019bcw} Photons, the quanta of the radiation potential, trap the particle and form the optical position sensor. Increasing the optical power raises the kick rate from individual photons, resulting in a force caused by shot noise of radiation pressure. As demonstrated in Ref.~\cite{Jain2016DirectMO}, the damping coefficient can be directly measured from the photon recoil during reheating.
\item The parametric feedback force $F_{\rm FB}$ and 
the corresponding damping coefficient $\gamma_{\rm FB}$. 
The feedback force are impressed on the contrary to the motion 
of levitated particle detected in the experiment.
Importantly, $F_{\rm FB}$ is a deterministic force that does not 
fluctuate, but rather damps the motion of the levitated particle. 
This is analogous to other damping effects. The main advancement 
of this paper lies in distinguishing between stochastic and
deterministic forces in thermal dynamics.
\item An additional stochastic force can be generated through 
collisions with the surrounding dark matter. Within the Langevin 
system, the interaction with dark matter is described by the 
damping coefficient $\gamma_{\rm DM}$ 
(We call it dark damping in the following.)
and the driving force $F_{\rm DM}$.
\end{enumerate}
It should be noted that laser trapping techniques generate a force where the driven force acts as the restoring force for the particle's oscillation. Furthermore, it is important to note that the damping coefficient $\gamma_{\rm RE }$ differs from $\gamma_{\rm FB}$ since they arise from different optical trapping setups. Dissipation and fluctuations can also occur due to self-emission, absorption, and scattering with ambient radiation from the environment. These dissipative and fluctuating effects are not considered in this paper. Naturally, the formulation in this section can be straightforwardly extended to include systems with other additional stochastic forces. It is important to keep in mind that the correlation function of the added stochastic force does not depend on a single uniform temperature.

We consider a scenario where a levitated particle with mass $m$ is created using a strongly focused laser beam of frequency $\Omega$ under ambient temperature and gas pressure conditions. In this setup, the particle is kicked by the incident photons from the environment, each carrying an energy of $\hbar \Omega$, resulting in the formation of a harmonic oscillator with a spring constant $k$ and a frequency $\omega_0=\sqrt{k/m}$. Typically, the frequency of the incident laser beam, $\Omega$, is significantly higher than the frequency of the levitation potential, $\omega_0$. The three spatial dimensions can be treated as decoupled from one another. Consequently, the motion of the Langevin system can be described as.
\begin{equation}
\frac{d^{2}q}{dt^{2}}+\gamma_{\rm T}\frac{dq}{dt}+\omega_{0}^{2}q=\frac{F_{\rm TH}\left(t\right)+F_{\rm RE}+F_{\rm DM}\left(t\right)+F_{\rm FB}\left(t\right)}{m}. \label{scho}
\end{equation}
The total damping coefficient $\gamma_{\rm T}$, represented by the second term on the left side of the above equation, accounts for the sum of damping coefficients arising from various sources
\begin{equation}
    \gamma_{\rm T} = \gamma_{\rm TH}+\gamma_{\rm RE}
    +\gamma_{\rm FB} +\gamma_{\rm DM}.
\end{equation}
The right-hand side represents the driven forces in the system. Generally, each damping coefficient corresponds to a term on the right-hand side representing the driven forces. However, as mentioned earlier, the feedback force in the laser cooling technique is deterministic. This force $F_{\rm FB}=-\omega_{0} \eta q^2 p$ is a time-varying, non-conservative damping optical force introduced through parametric feedback with a strength $\eta$, where $p=m \dot{q}$ represents the momentum.

In Langevin systems, the average value of any stochastic force $F_i$ is zero. The correlation functions should follow a similar form as Eq.~\eqref{general-core}, which is proportional to the damping coefficients and temperature if they satisfy the fluctuation dissipation theorem. Here, $T_i$ does not represent a uniform temperature but should be regarded as an effective parameter that characterizes the correlation function of each individual stochastic force. Naturally, $T_{\rm TH}$ corresponds to the temperature of the ambient environment $T_{\rm E}$. The sum of all stochastic forces is denoted as.
\begin{equation}
F_{\rm S}(t)=F_{\rm TH}\left(t\right)
+F_{\rm RE}\left(t\right)
+F_{\rm DM}\left(t\right)\, .
\end{equation}
It is evident that the equilibrium of the steady state of the levitated particle does not imply that its temperature is equal to that of the environment. A new stochastic temperature, denoted as $T_{\rm S}$, can be defined as
\begin{equation}
T_{\rm S}=\frac{\gamma_{\rm TH}T_{\rm TH}+\gamma_{\rm RE}T_{\rm RE}+\gamma_{\rm DM}T_{\rm DM}}{\gamma_{\rm TH}+\gamma_{\rm RE}+\gamma_{\rm DM}}\, ,
\label{defTs}
\end{equation}
which accounts for the temperature of the system in the correlation function
\begin{equation}
\langle F_{\rm S}(t) F_{\rm S}(t') \rangle = 2m \gamma_{\rm S} k_{\rm B} T_{\rm S} \delta (t-t')\, .
\label{core-F}
\end{equation}
In the above equation, $\gamma_{\rm S}$ represents the sum of all damping coefficients associated with the respective stochastic forces
\begin{equation}
\gamma_{\rm S}=\gamma_{\rm TH}+\gamma_{\rm RE}+\gamma_{\rm DM}\, .
\label{defgs}
\end{equation}
It is important to note that the system temperature $T_{\rm S}$ does not describe the thermal state of the Langevin system since the levitated particle represents a single degree of freedom. The parameter $T_{\rm S}$, along with all the $T_i$ in Eq.~\eqref{defTs}, characterizes the nature of the stochastic forces. In the absence of a feedback loop, $T_{\rm S}$ represents the temperature of the environment in an ordinary Langevin system. In experiments, the dynamical relaxation can be measured by manipulating the feedback loop switch.~\cite{Jain2016DirectMO} The heating or cooling trajectories can provide detailed information about the stochastic forces. As mentioned in the introduction, the Langevin system contains multi-stochastic forces from different sources. The phenomenology of heating or cooling processes may deviate from predictions if only two parameters, $\gamma_{\rm T}$ and $T_{\rm S}$, are used to account for all the stochastic forces. Therefore, the following section discusses parametric feedback cooling with multiple stochastic forces.

As demonstrated in Ref.~\cite{2014NatNa...9..358G}, 
the energy of the Langevin system can be expressed in a form that resembles under-damped Brownian motion in the energy space. Expressing 
Eq.~\eqref{scho} as a stochastic differential equation (SDE) is more convenient.
\begin{eqnarray}
{\rm d} q & =& \frac{p}{m} {\rm d} t \label{eqdq}\,, \nonumber \\
{\rm d}p&=&\left(-m\omega_0^{2}q -\omega_0\eta q^2 p -\gamma_{\rm TH}p
\right.\nonumber \\
& & \left. -\gamma_{\rm RE}p
 -\gamma_{\rm FB}p-\gamma_{\rm DM}p\right){\rm d }t\\
& &  +\sqrt{2m\gamma_{\rm TH}k_{\rm B}T_{\rm TH}}{\rm d }W_{\rm TH }\nonumber \\
& & +\sqrt{2m\gamma_{\rm RE}k_{\rm B}T_{\rm RE}}{\rm d }W_{\rm RE }\nonumber \\
& &  +\sqrt{2m\gamma_{\rm DM}k_{\rm B}T_{\rm DM}}{\rm d }W_{\rm DM}\,,\nonumber 
\end{eqnarray}
where $W_i(t)$ are Wiener processes with 
\begin{equation}
\langle W_i(t)\rangle  =0,~~~~
\left\langle W_i(t) W_j
\left(t^{\prime}\right)\right\rangle =\delta_{ij}(t^{\prime}-t)\,.
\end{equation}
Here only the thermal and additional dark  stochastic forces
are taken into account. 
For a short (infinitesimal) time interval $\mathrm{d} t$ we have
\begin{equation}
\langle\mathrm{d} W_i\rangle =0,~~~~ 
\left\langle(\mathrm{d}W_i)^2\right\rangle  =\mathrm{d} t .
\end{equation}
Next, the energy change $\mathrm{d} E$ is determined by
\begin{equation}
\mathrm{d} E=\left(\frac{\partial E}{\partial q}\right) \mathrm{d} q+\left(\frac{\partial E}{\partial p}\right) \mathrm{d} p+\frac{1}{2}\left(\frac{\partial^2 E}{\partial p^2}\right)(\mathrm{d} p)^2\, .
\end{equation}
It is important to note that the Wiener process implies that $(\mathrm{d} p)^2$ is of the same order as $\mathrm{d} t$. Therefore, the second-order partial differential of the energy $E$ needs to be considered, while $(\mathrm{d} q)^2$ and $\mathrm{d} q \mathrm{d} p$ can be neglected. By taking the derivatives of the energy with respect to $q$ and $p$, we obtain
\begin{equation}
\mathrm{d} E=m \omega_0^2 q \mathrm{d} q +\frac{p}{m} \mathrm{d} p+\frac{1}{2 m}(\mathrm{d} p)^2\,.
\end{equation}
Inserting $\mathrm{d} q,~\mathrm{d} p$ and neglecting all terms of order 
$(\mathrm{d} t)^{3 / 2}$ or higher yields
\begin{eqnarray}
\mathrm{d} E &=& \left(-\gamma_{\rm T} \frac{p^2}{m} 
-\frac{\eta E^2}{2 m \omega_0}\right) \mathrm{d} t
+\frac{\gamma_{\rm S} k_{\mathrm{B}} T_{\rm S}}{2}
\mathrm{d} t\label{eqde2}\nonumber \\
&& +\sqrt{2 E \gamma_{\rm TH} k_{\mathrm{B}} T_{\rm TH}} 
\mathrm{d} W_{\rm TH} \\
& &+\sqrt{2 E \gamma_{\rm RE} k_{\mathrm{B}} T_{\rm RE}} 
\mathrm{d} W_{\rm RE} \nonumber \\
& & +\sqrt{2 E \gamma_{\rm DM} k_{\mathrm{B}} 
T_{\rm DM}} \mathrm{d} W_{\rm DM}\,.\nonumber
\end{eqnarray}
Note that an additional factor of $1/2$ arises in the second term on the right-hand side when performing stochastic integration over the time evolution of the harmonic oscillation. This formula elucidates the fundamental distinction between energy increments driven by single and multiple stochastic forces. For a single stochastic force, the formulation comprises a single damping coefficient and its corresponding temperature. However, the presence of multiple stochastic forces considerably complicates the formulation, impacting the energy distribution, dissipation, relaxation, and other factors. Each term $\gamma_i k_{\rm B}T_i{\rm d}t$ arises in association with a corresponding stochastic force. However, additional contributions to the damping terms, $\gamma_i p^2 {\rm d}t/m$, arise from the feedback mechanism. This additional damping force can, in fact, be any dissipative force that does not involve fluctuations to drive the levitated particle. Consequently, the total damping coefficient $\gamma_{\rm T}$, which incorporates the additional contribution of $\gamma_{\rm FB}$, is not equal to the stochastic damping coefficient $\gamma_{\rm S}$. As demonstrated below, this difference in damping coefficients gives rise to distinct thermal dynamics compared to those resulting from a single stochastic force, even when $T_{\rm S}$ is erroneously assumed to be the environmental temperature $T_{\rm E}$. 

\begin{figure*}[htpb]
\includegraphics[width=54mm]{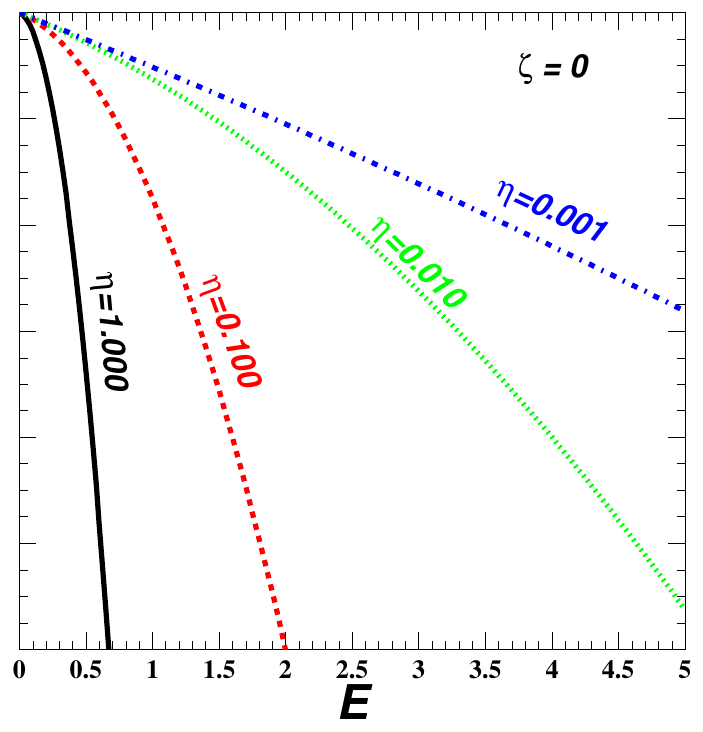}
\includegraphics[width=54mm]{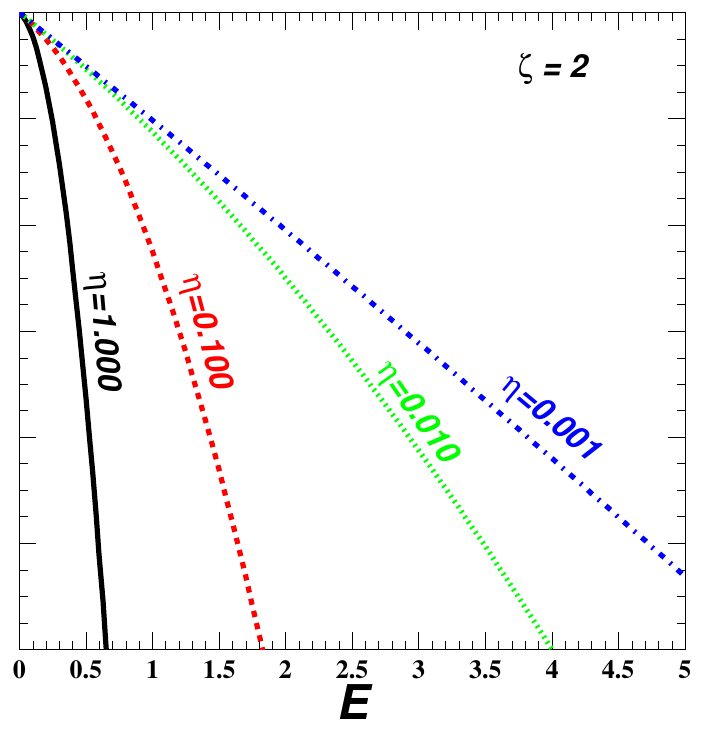}
\includegraphics[width=54mm]{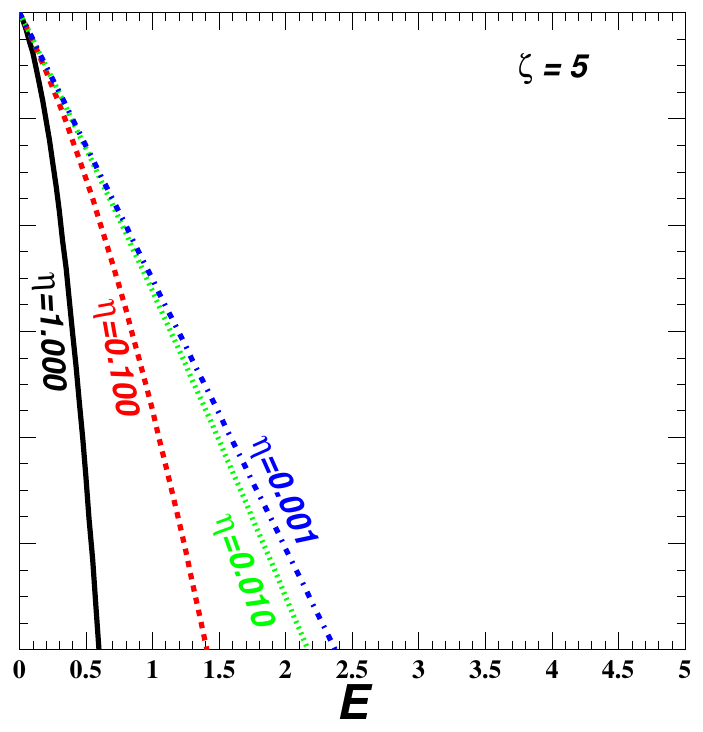}
\caption{The logarithm of the energy distribution $\rho(E, \alpha, ~\zeta)$ is compared for different $\zeta$ values. The distributions are normalized with $\ln\rho(0, \alpha, ~\zeta)=0$. The parameters are set as $k_{\rm S}T_{S}=1$, $m=1$, $k=1$, and $\gamma_{\rm S}=0.01$. Each curve corresponds to a different feedback strength $\eta$ shown in the plot. The plots for $\zeta=2$ and $\zeta=5$ depict the distributions with an additional damping factor.}
\label{sec2-figpe}
\end{figure*}
Following a similar procedure outlined in the appendix of 
Ref.~\cite{2014NatNa...9..358G}, we adopt $\epsilon = \sqrt{E}$ to study the energy increment and prevent the emergence of multiplicative noise in Eq.~\eqref{eqde2}
\begin{eqnarray}
\mathrm{d}\epsilon&=&\left(-\frac{\gamma_{\rm T}\epsilon}{2}
-\frac{\eta\epsilon^{3}}{4m\Omega_{0}}+\frac{\gamma_{\rm S}
k_{\mathrm{B}}T_{{\rm S}}}{4\epsilon}\right)\mathrm{d}t\\
&& +\sqrt{\frac{\gamma_{{\rm TH}}k_{\mathrm{B}}T_{\rm TH}}{2}}\mathrm{d}W_{{\rm TH}}
+\sqrt{\frac{\gamma_{{\rm RE}}k_{\mathrm{B}}T_{\rm RE}}{2}}\mathrm{d}W_{{\rm RE}}\nonumber\\
&& +\sqrt{\frac{\gamma_{{\rm DM}}k_{\mathrm{B}}T_{{\rm DM}}}{2}}\mathrm{d}W_{{\rm DM}}\,.\nonumber
\end{eqnarray}
The variable $\epsilon$ evolves at temperature $T_{\rm S}$ under the influence of an external force $f(\epsilon)$ in the presence of high friction characterized by $\nu=4/\gamma_{\rm S}$
\begin{eqnarray}
    \mathrm{d}\epsilon&=&\frac{1}{\nu }f(\epsilon) {\rm d}t
    +\sqrt{\frac{\gamma_{{\rm TH}}k_{\mathrm{B}}T_{\rm TH}}{2}}\mathrm{d}W_{{\rm TH}}\\
    && +\sqrt{\frac{\gamma_{{\rm RE}}k_{\mathrm{B}}T_{\rm RE}}{2}}\mathrm{d}W_{{\rm RE}}+\sqrt{\frac{\gamma_{{\rm DM}}k_{\mathrm{B}}T_{{\rm DM}}}{2}}\mathrm{d}W_{{\rm DM}}\, .\nonumber
\end{eqnarray}
We can denote a new parameter  
\begin{equation}
    \zeta=\frac{\gamma_{\rm T}}{\gamma_{\rm S}}
\end{equation}
to account for the 
difference between stochastic damping and the total damping.
Then the external force 
\begin{eqnarray}
f\left(\epsilon\right)&=&-2\zeta\epsilon-\frac{\eta\epsilon^{3}}{m\omega_{0}\gamma_{\rm S}}+\frac{k_{\mathrm{B}}T_{{\rm S}}}{\epsilon}\nonumber\\
&=&-\frac{{\rm d}U\left(\epsilon\right)}{{\rm d}\epsilon}\,.    
\end{eqnarray}
Then the potential $U\left(\epsilon\right)$ should be
\begin{equation}
U\left(\epsilon\right)=\zeta \epsilon^{2}+\frac{\eta}{4m\omega_{0}\gamma_{\rm S}}\epsilon^{4}-k_{\mathrm{B}}T_{{\rm E}}\ln\epsilon\,.
\end{equation}
Denote
\begin{equation}
    \alpha =\frac{\eta}{m\omega_{0}\gamma_{\rm S}},~~
    \beta_{\rm S}=\frac{1}{k_{\mathrm{B}}T_{{\rm S}}}.
\end{equation}
Then the ergodic hypothesis in statistical physics
can be used, and 
the distribution of $\epsilon$ can be easily derived 
by Boltzmann-Gibbs distribution
\begin{equation}
\rho(\epsilon,~\alpha, ~\zeta) \propto \exp \left\{-\frac {U(\epsilon)}{k_{\mathrm{B}}T_{{\rm S}}}\right\} \propto \epsilon \exp \left\{-\beta_{\rm{S}}\left(\zeta \epsilon^2+\frac{\alpha}{4} \epsilon^4\right)\right\}.\label{rhoEalpha}
\end{equation}
The distribution function of energy can be got by 
recovering  variables from $\epsilon$ to $E=\epsilon^2$
\begin{equation}
\rho(E, \alpha, ~\zeta)=\frac{1}{Z_{\alpha\zeta }}
\exp \left\{-\beta_{\rm{S}}
\zeta E-\beta_{\rm{S}}\frac{ \alpha}{4} E^2\right\},
\label{rhoEalpha1}
\end{equation}
where the normalisation factor 
$Z_{\alpha\zeta }=\int \mathrm{d} E \rho(E, \alpha)$ is given by
\begin{equation}
Z_{\alpha\zeta }=\sqrt{\frac{\pi}{\alpha \beta_{\rm{S}}}} e^{\zeta^2 \beta_{\rm{S}} / \alpha} \operatorname{Erfc}\left(\sqrt{ \frac{\zeta^2\beta_{\rm{S}}}{\alpha}}\right).
\end{equation}
The distribution clearly reverts to the form of the single stochastic force when $\zeta = 1$. However, the temperature $T_{\rm S}$ remains determined by multiplicative forces. To demonstrate the modification of the energy distribution, we numerically calculated the logarithm of the distribution $\rho(E, \alpha, \zeta)$ for different $\zeta$ values. The results are depicted in Fig.~\ref{sec2-figpe}. From the figure, it is evident that increasing the $\zeta$ parameter sharpens the distribution as the energy increases. This can be easily understood as a consequence of the additional damping factor ($\gamma_{\rm FB}$) leading to increased energy dissipation.

The modification of the distribution can be observed through the average energy or the effective temperature of the levitated particle. The average energy is obtained by integrating over the energy distribution given by Eq.~\eqref{rhoEalpha1}.
\begin{equation}
\langle E\rangle=Z_\alpha^{-1} \int \mathrm{d} E E \exp \left\{-\beta_{\rm{S}} \zeta E-\frac{\beta_{\rm{E}} \alpha}{4} E^2\right\}\,.
\end{equation}
Evaluation of the integral yields
\begin{equation}
\langle E\rangle=\frac{2}{\beta_{\rm{S}}}\left(\sqrt{\frac{\beta_{\rm{S}}}{\alpha}} \frac{e^{-\zeta^2 \beta_{\rm{S}} / \alpha}}{\sqrt{\pi} \operatorname{Erfc}\left(\sqrt{\frac{\zeta^2 \beta_{\rm{S}}}{\alpha}}\right)}-\frac{\zeta \beta_{\rm{S}}}{\alpha}\right)\, .
\end{equation}
If defined an effective temperature obtained by 
applying the feedback mechanism
$\langle E\rangle=k_{\rm B} T_{\text{eff}}$, then 
\begin{equation}
 T_{\text{eff}}= T_{\rm{S}}\left\{\frac{2}{\sqrt{\alpha k_{\mathrm{B}} T_{\rm{S}}}} \frac{e^{-\zeta^2 / \alpha k_{\mathrm{B}} T_{\rm{S}}}}{\sqrt{\pi} \operatorname{Erfc}\left(\frac{\zeta}{\sqrt{\alpha k_{\mathrm{B}} T_{\rm{S}}}}\right)}-\frac{2\zeta}{\alpha k_{\mathrm{B}} T_{\rm{S}}}\right\}.
\end{equation}

\begin{figure}[htpb]
\begin{center}
\includegraphics[width=70mm]{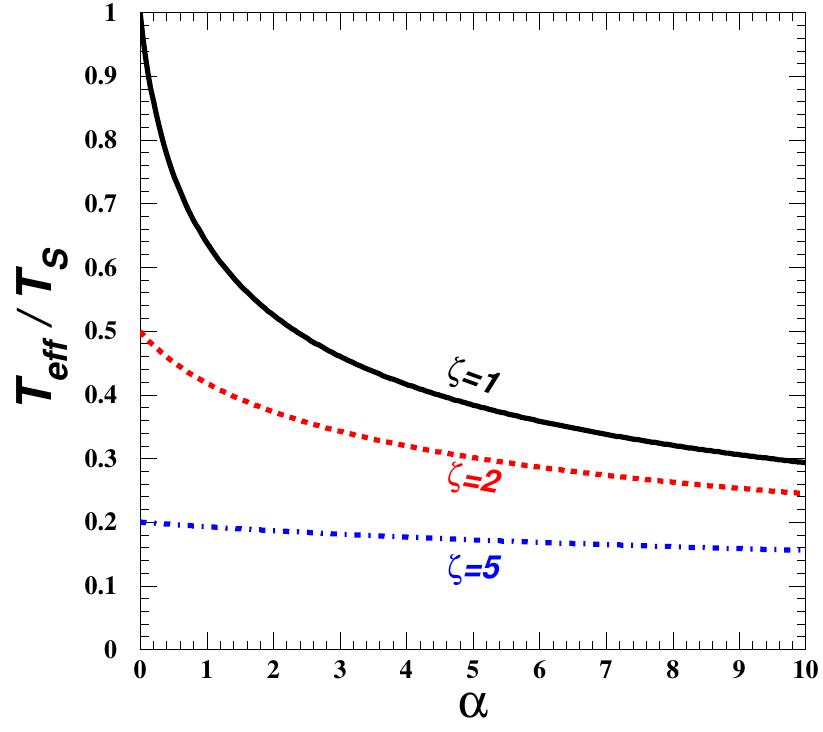}
\caption{ The ratio of the effective temperature 
to the stochastic temperature $T_{\text{eff}}/T_{\rm S}$
versus $\alpha$. $\zeta=2$ and $\zeta=5$ curves
show the depression of effective temperature 
with additional damping factor.}\label{sec2-figefft}
\end{center}
\end{figure}

Fig.~\eqref{sec2-figefft} shows the numerical results of the ratio $T_{\text{eff}}/T_{\rm S}$ as a function of the parameter $\alpha$. It can be observed that the effective temperature decreases with increasing $\alpha$. Moreover, the effective temperature is further reduced by the additional damping factor introduced by the feedback mechanism. This yields a highly interesting result.
\begin{equation}
    \lim_{\alpha\to 0} \frac{T_{\rm eff}}{T_{\rm S}}=\frac{1}{\zeta}\,.
    \label{limteffts}
\end{equation}
When $\alpha\to 0$, it signifies the absence of the driven force from the feedback. Hence, the term $1/\zeta$ quantifies the dampening effect of the additional damping factor. This mechanism effectively cools the levitated particle by interacting with the environment. Additionally, a non-zero feedback strength enhances the cooling effect.

The averages over the last two terms of Eq.~\eqref{eqde2} vanish
$ \langle\mathrm{d} W_{\rm TH}\rangle =0,~   
 \langle\mathrm{d} W_{\rm DM}\rangle =0$.
Thus the time evolution of the average energy is 
\begin{eqnarray}
\frac{\mathrm{d} \langle E\rangle}{\mathrm{d} t}&=& -\gamma_{\rm T} 
\langle E\rangle - \frac{\eta}{2 m \omega_0}\langle E^2 \rangle 
+\frac{\gamma_{\rm S} k_{\mathrm{B}} T_{\rm S}}{2}\label{eqavde}\\
&=& -\gamma_{\rm T} 
\langle E\rangle - \frac{\gamma_{\rm S}\alpha }{2 }\langle E^2 \rangle 
+\frac{\gamma_{\rm S} k_{\mathrm{B}} T_{\rm S}}{2}\, .\nonumber     
\end{eqnarray}
This equation describes the cooling process in which the average energy decreases with the evolution due to the presence of feedback. In the steady state, the time derivative on the left side of Eq.~\eqref{eqavde} becomes zero.
\begin{equation}
    \alpha \langle E^2\rangle=k_{\rm B}T_{\rm S}
    -2\zeta \langle E\rangle\, .
\end{equation}
The average dissipation rate $\bar P$ of 
the feedback mechanism can be formulated as
\begin{equation}
    \bar P = \gamma_{\rm FB} \langle E\rangle
    +\frac{\gamma_{\rm S}\alpha \langle E^2\rangle}{2}\, .
\end{equation}
Since the second moment of the energy $\langle E^2 \rangle$ in Eq.~\eqref{eqavde} cannot be expressed in terms of the average energy $\langle E \rangle$, the equation cannot be easily solved for $\langle E \rangle$. However, in the equilibrium state for the exponential distribution, we have $\langle E^2 \rangle=2\langle E \rangle^2$. Therefore, the cooling process can be approximately derived.
\begin{widetext}
\begin{equation}
    \langle E(t)\rangle =
    \frac{\zeta}{2\alpha}\frac{-1+\sqrt{1+2\alpha k_{\rm B}T_{\rm S}/\zeta^{2}}+\left(1+\sqrt{1+2\alpha k_{\rm B}T_{\rm S}/\zeta^{2}}\right)C\exp\left(-\gamma_{\rm T}\sqrt{1+2\alpha k_{\rm B}T_{\rm S}/\zeta^{2}}t\right)}{1-C\exp\left(-\gamma_{\rm T}\sqrt{1+2\alpha k_{\rm B}T_{\rm S}/\zeta^{2}}t\right)}\,.
    \label{eqcooling}
\end{equation}
\end{widetext}
Here, $C$ is an arbitrary constant determined by the initial condition. It is reasonable to assume that the cooling process initiates when the levitated particle reaches a steady state with the ambient stochastic forces in the absence of feedback. In the levitation experiment, the feedback force can be toggled on or off, allowing for the measurement of both the cooling and heating processes of the levitated particle. The heating process is conducted by turning off the feedback loop, resulting in the evolution of the average energy according to
\begin{equation}
\frac{\mathrm{d}\langle E\rangle}{\mathrm{d} t}=-\gamma_{\rm S}\langle E\rangle+\frac{\gamma_{\rm S} k_{\mathrm{B}}T_{\rm S}}{2}\, .
\label{eqdedt}
\end{equation}
It is important to note that when the feedback is turned off, the damping and fluctuation arise from the collective ambient stochastic forces discussed earlier. Therefore, in Eq.~\eqref{eqavde}, $\gamma_{\rm T}$ must be replaced by $\gamma_{\rm S}$. However, the presence of multiple sources implies that the system temperature does not necessarily equal the environmental temperature. The relaxation evolution can be easily solved
\begin{equation}
\begin{aligned}
\langle E(t)\rangle &=\frac{k_{\mathrm{B}}T_{\rm S}}{2}+\left[\langle E(0)\rangle-\frac{k_{\mathrm{B}}T_{\rm S}}{2}\right]
\mathrm{e}^{-\gamma_{\rm S} t}\\
&=\frac{k_{\mathrm{B}}T_{\rm S}}{2}+\left[\frac{k_{\rm B} T_{\rm L}}{2}
-\frac{k_{\mathrm{B}}T_{\rm S}}{2}\right]
\mathrm{e}^{-\gamma_{\rm S} t}\,.\label{eqheating}
\end{aligned}
\end{equation}
The temperature $T_{\rm L}$ in the above equation represents the newly introduced effective temperature that characterizes the initial heating state. By examining Eq.~\eqref{eqheating}, it becomes apparent that as time progresses, the levitated particle will tend to reach the ambient temperature $T_{\rm H} = T_{\rm S}$. This result can also be obtained from Eq.~\eqref{eqdedt}, where the time derivative on the left-hand side vanishes in the equilibrium state. At this point, the energy flow from fluctuations compensates for the dissipation.

If the feedback is turned on, the system will be cooled and will no longer be in equilibrium with the ambient stochastic forces. By requiring $\langle E(0)\rangle =k_{\mathrm{B}}T_{\rm H}/2$, we obtain

\begin{equation}
    C=\frac{\alpha k_{\rm B}T_{\rm H}+\zeta-\sqrt{\zeta^{2}+2\alpha k_{\rm B}T_{\rm H}}}{\alpha k_{\rm B}T_{\rm H}+\zeta+\sqrt{\zeta^{2}+2\alpha k_{\rm B}T_{\rm H}}}\,.
\end{equation}
It is straightforward to determine that $0 < C < 1$ based on the condition $\zeta > 1$. By examining Eq.~\eqref{eqcooling}, we can determine the final steady state of the cooling process, where the energy flow from the oscillator to the stochastic source is counterbalanced by the energy extracted from the feedback mechanism
\begin{equation}
    \langle E(t \to \infty )\rangle =\frac{\zeta}{2\alpha}\left(\sqrt{1+\frac{2\alpha k_{\rm B}T_{\rm H}}{\zeta^{2}}}-1\right)\,.
\end{equation}
Therefore, the approximate effective temperature at the cooling limit should be.
\begin{equation}
    T_{\rm L}^{\rm limit} =
    \frac{\zeta}{\alpha k_{\rm B}}\left(\sqrt{1+\frac{2\alpha k_{\rm B}T_{\rm H}}{\zeta^{2}}}-1\right)\,.\label{coollim}
\end{equation}
It can be observed that the ratio $\beta\equiv {2\alpha k_{\rm B}T_{\rm H}}/{\zeta^{2}}$ determines the simple expression for the cooling limits. If $\beta \ll 1$, the cooling limit is expected to be
\begin{equation}
    T_{\rm L}^{\rm limit} \simeq \frac{T_{\rm H}}{\zeta}\,.\label{coollimdamp}
\end{equation}
This indicates that the cooling process is dominated by the feedback damping $\gamma_{\rm FB}$. The cooling limit is in agreement with the effective temperature given by Eq.~\eqref{limteffts}. However, in the case of $\beta \gg 1$,
\begin{equation}
    T_{\rm L}^{\rm limit} \simeq \sqrt{\frac{2T_{\rm H}}{\alpha k_{\rm B}}}\,,\label{coollimopt}
\end{equation}
This is independent of $\gamma_{\rm FB}$. This implies that the cooling process is dominated by the feedback optical force, and it also necessitates ${2}/{\alpha k_{\rm B}}<T_{\rm H}$. Otherwise, cooling will not be achieved. These limits differ from the results in Ref.~\cite{2014NatNa...9..358G}, which are solely determined by the balance between single fluctuation and feedback optical force.

\begin{figure}[htpb]
\begin{center}
\includegraphics[width=68mm]{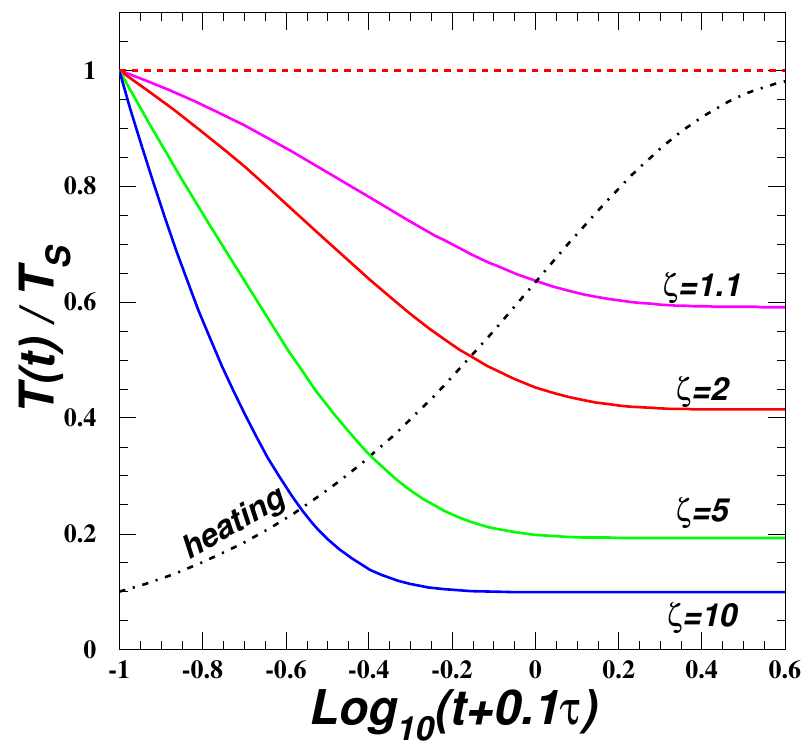}
\caption{The numerical results for the heating and cooling processes are presented. The parameters $\alpha k_{\rm B}$ and $\gamma_{\rm S}$ are normalized to unity in the numerical calculation. The dashed line represents the heating process, which evolves exponentially as $\exp(-\gamma_{\rm S}t)$ with time. The solid lines represent the cooling processes for different values of the parameter $\zeta$. The time evolution is scaled by a factor of $0.1\tau$ to enable a logarithmic representation.}\label{sec2-col}
\end{center}
\end{figure}

Fig.~\ref{sec2-col} illustrates the numerical results for the heating and cooling processes. Parameters $\alpha k_{\rm B}$ and $\gamma_{\rm S}$ are normalized to unity in the numerical calculation, and the ratio of the effective temperature to the ambient temperature $T_{\rm S}$ is presented. The heating process evolves exponentially as $\exp(-\gamma_{\rm S}t)$ with time, while the cooling rate is determined by the parameter $\zeta$ and evolves as $\exp(-\gamma_{\rm T}\sqrt{1+2\alpha k_{\rm B}T_{\rm S}/\zeta^{2}}t)$. A larger value of $\zeta$ implies faster cooling. Here, it is important to address the differences between our work and the theory proposed in Ref.~\cite{2014NatNa...9..358G}. In our work, cooling can be achieved through both feedback damping and feedback optical force. When $\gamma_{\rm FB}\gg \gamma_{\rm S}$, the feedback damping force dominates the dissipation process. However, in the limit $\gamma_{\rm FB}\to 0$ or equivalently $\zeta\to 1$, the cooling process is the same as the one proposed in Ref.~\cite{2014NatNa...9..358G}, where the optical force cools the system. In the following section, we will discuss the distinctions of the several damping forces.


\section{ The damping forces from different sources}\label{sec3-st}

Based on the thermal dynamics presented in the previous section, it is evident that the damping factor and temperature are the two fundamental parameters for describing properties, such as correlation functions of the stochastic forces. The damping coefficients can be viewed as macroscopic phenomena that arise from collective interactions among the microscopic particles comprising both the levitated particle and the environment. These interactions can be attributed to mechanical or electromagnetic dynamics in the microscopic realm. To gain a comprehensive understanding of the physics of levitation and enhance the sensor's sensitivity, it is crucial to thoroughly investigate the collective interactions. This will allow us to modify experimental setups and unveil the underlying mechanism. Therefore, in this section, we explore the microscopic origins of the three damping coefficients discussed in the previous section.   
Since $\gamma_{\rm TH}$, $\gamma_{\rm RE}$, and $\gamma_{\rm FB}$ have been extensively studied in the literature, and this paper aims to explore new methods for dark matter detection, we will provide a brief introduction to $\gamma_{\rm TH}$, $\gamma_{\rm RE}$, and $\gamma_{\rm FB}$. The formulation of $\gamma_{\rm DM}$ and the distinctions between this coefficient and the other three will be discussed in detail.

\subsection{The isothermal drag force}
\begin{figure*}[htpb]
\begin{center}
\includegraphics[width=160mm]{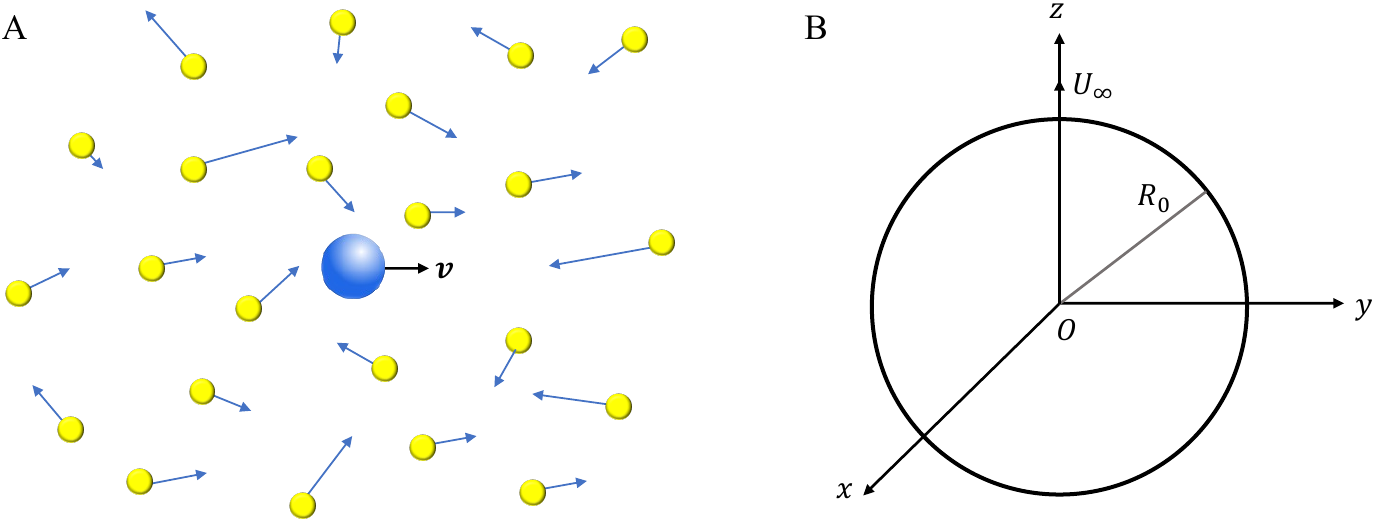}
\caption{(A) Figure illustrating the Knudsen number in fluid dynamics: a stationary flow past a particle in its own saturated vapor. The damping is determined by the mean free path of the molecules and the length scale of the particle.
(B) Calculation scheme for isothermal drag: a spherical ball with a radius $R_0$ moving through the fluid with a vapor velocity of $U_\infty$.}\label{sec3-kn}
\end{center}
\end{figure*}
The first damping rate we need to consider is the isothermal drag force, $\gamma_{\rm TH}$, which arises from the interaction between a spherical particle and rarefied gas molecules. A single aerosol particle suspended in a non-equilibrium gas can experience various forces due to non-uniformity, including isothermal drag, thermal force, photophoretic force, diffusion force, and other forces resulting from combined flows of heat, mass, and momentum. Detailed investigations of these forces are beyond the scope of this paper, but they can be found in standard fluid dynamics textbooks.~\cite{tritton2012physical} Clearly, the isothermal drag force is the primary contributor to the damping force experienced by the levitated particle in the fluid. As depicted in panel A of Fig.~\ref{sec3-kn}, a particle in its own saturated vapor experiences a stationary flow. The physical phenomena are attributed to phase changes occurring on the particle surface.
The theoretical approach to the damping coefficient
depends on the so called Knudsen number, denoted as $Kn$
\begin{equation}
    Kn=\frac{\lambda }{L}\,.
\end{equation}
Here, $\lambda$ represents the mean free path of the microscopic component particles in the surrounding matter, while $L$ refers to the characteristic length scale of the levitated particle.
The schematic diagram can be observed in plot B of Fig.~\ref{sec3-kn}, illustrating a spherical particle with radius $R_0$ surrounded by a rarefied gas consisting of molecules with mass $m_{M}$. The number density of molecules is denoted as $n_\infty$, while the vapor velocity is denoted as $U_\infty$. 
To derive the damping rate, we need to do 
the linearization of the vapor velocity $U_\infty$ under the
assumption that $U_\infty$ is significantly smaller than the velocity of the molecules, resulting in a ratio $u_\infty$.
\begin{equation}
    u_\infty = U_\infty \left(\frac{m_{M}}{2k_{\rm B} T_{\rm TH}}\right)
    \ll 1 \, .\label{eqlincon}
\end{equation} 
is denoted and the molecular distribution can be linearized as 
\begin{equation}
f(\bm{r}, \bm{v}) = f_{\infty}\left[ 1 + 2 \bm{c} \cdot \bm{u}_{\infty}+\cdots\right]\, ,
\label{eqlindis}
\end{equation}
where $f_{\infty}$ is the Maxwell-Boltzmann distribution. It is important to emphasize that the linearization of the velocity distribution is a crucial step in the analysis. However, when the levitated particle collides with the surrounding dark matter, the condition given by Eq.~\eqref{eqlincon} fails, and the linearization operation is not applicable. The temperature $T_{\rm TH}$ in Eq.~\eqref{eqlincon} has the same meaning as described in the previous section, referring to the temperature of the heating bath.

A comprehensive investigation of the isothermal drag force can be found in Ref.~\cite{beresnev_chernyak_fomyagin_1990}. Here, we provide a summary of the coefficient results for different cases as follows. When $Kn\ll 1$, the molecules in the surrounding matter can be treated as material fragments of the moving medium. This region corresponds to the fundamental concept of continuum description in fluid flows, known as the viscous slip-flow regime. The analytical expression for the damping coefficients can be written as
\begin{eqnarray}
 \gamma_{\rm TH}&=&\frac{F^{\rm V}}{m U_\infty}\\
    &=& \frac{6\pi}{m} \eta_{\rm g} R_0 (1+a Kn
    +b Kn^2 +{\cal O} (Kn^3))\,,\nonumber 
\end{eqnarray}
where $Kn= \lambda /R_0$.  $\eta_{\rm g}$ represents the gas viscosity at temperature $T_{\rm TH}$, while $a$ and $b$ are numerical coefficients that depend on the momentum accommodation during collisions between the molecules and the spherical particle. The damping coefficient can be related to the Stokes formula used in statistical physics. When $Kn\gg 1$, the surrounding molecules can be treated as freely moving, leading to what is known as the free-molecular regime. In this regime, the damping coefficients are
\begin{eqnarray}
\gamma_{\rm TH}&=&\frac{F^{\rm F}}{m U_\infty}\label{gamthpg}\\
&=&\frac{8\pi^{\frac{1}{3}}}{3m}  R_0^2 P_{\rm TH} \left(\frac{m_M}{2 k_{\rm B} T_{\rm TH}}\right)^{\frac{1}{2}}\nonumber \\
&&\times\left\{2+\alpha_\tau-\alpha_n \frac{32-\pi\left(9-\alpha_E\right)}{32-\pi\left(1-\alpha_n\right)\left(9-\alpha_E\right)}\right\}\,,\nonumber
\end{eqnarray}
where $P_{\text {TH}}$ denotes the pressure of the surrounding gas. The momentum and energy accommodation coefficients $\alpha_r,\alpha_n,\alpha_E$ can be found in tables listed in Ref.~\cite{beresnev_chernyak_fomyagin_1990} and the accompanying references. It can be observed that adjusting the pressure leads to different thermal damping coefficients. Since nanoparticle levitation experiments are typically conducted in ultra-high vacuum, this property can be utilized to mitigate or eliminate thermal damping and fluctuations in the levitation, as demonstrated in the next section.

\subsection{The photon recoil in the optical levitation}
The trapping potential in the levitation is achieved through the use of a strongly focused laser, which imparts a photon kick to the levitated particle. Subsequently, fluctuations occur due to classical noise in the laser intensity, causing modulation of the trapping potential. Additionally, a damping force arises during collisions to prevent the energy of the levitated particle from diverging. The damping coefficient can be deduced from the fluctuations of the trapping potential. The Hamiltonian for the trapped particle is given by ~\cite{PhysRevA.58.3914}
\begin{equation}
    H= \frac{p^2 }{2m}+\frac{1}{2}m\omega_0^2 (1+\epsilon(t))q^2\, ,
    \label{harosc}
\end{equation}
where  $\epsilon(t)$ represents the newly introduced time-dependent factor that accounts for the fluctuation of the laser intensity. The rate is reduced by the ratio of the parametric resonance linewidth, $\epsilon_0\omega_0$, to the bandwidth of the fluctuations, $\Delta \omega_0$. Here, $\epsilon_0$ denotes the root-mean-square fractional fluctuation in the spring constant $k$. Consequently, the rate scales as $\omega_0^2 S$, with $S\simeq \epsilon_0^2/\Delta \omega_0$ representing the noise spectral density in fractions squared per rad/sec. 

The damping rate can be computed classically, which is consistent with expectations for a harmonic-oscillator potential. First-order time-dependent perturbation theory can be employed to calculate the average transition rates between quantum states of the trap. The time-evolving perturbation of the quantum-mechanical Hamiltonian given by Eq.~\eqref{harosc} is
 \begin{equation}
     H'(t)=\frac{1}{2}\epsilon(t)m \omega_0^2 q^2\,.
     \label{hprem}
 \end{equation}
The damping of the levitated particle can be understood as the transition from a higher energy level $|n\rangle$ to a lower level $|m\rangle$ induced by the perturbation. The rate can be calculated by taking the average over a time interval $T$
\begin{eqnarray}
     &&R_{n\to m}=\frac{1}{T}\left|
     \frac{-i}{\hbar}\int_0^T{\rm d}tH'_{mn}(t)e^{i\omega_{mn}t}\right|^2\label{trans}\\
    & &= \left(\frac{m\omega_0^2}{2\hbar}\right)^2\nonumber
    \int_{-\infty }^{\infty }{\rm d}\tau
     e^{i\omega_{mn}\tau}\langle\epsilon(t)\epsilon(t+\tau)\rangle
     \left|\langle m|q^2|n\rangle\right|^2\,.\nonumber         
\end{eqnarray}
Here, we assume that the averaging time $T$ is short compared to the time scale over which the level populations vary, but large compared to the correlation time of the fluctuations. This allows the range of $\tau$ to formally extend to $\pm\infty$.
 \begin{equation}
     \langle\epsilon(t)\epsilon(t+\tau)\rangle
     =\frac{1}{T}\int_0^T {\rm d}t \epsilon(t)\epsilon(t+\tau)\,.
 \end{equation}
 
 Using the transition matrix elements of $q^2$ and 
 $\omega_{n\pm 2, n}$ in Eq.~\eqref{trans}, we can get the
 transition rate 
 \begin{equation}
R_{n\pm 2\to n}=\frac{\pi^2 \omega_0^2 }{16} S_k(2\omega_0 )
(n+1\pm 1)(n\pm 1 )\,, \label{rn2to2}
 \end{equation}
 in which $S_k $ is the one-sided power spectrum of the fractional 
 fluctuation in the spring constant
 \begin{equation}
     S_k(\omega_0)=\frac{2}{\pi }\int_0^\infty{\rm d}\tau
     \cos(\omega_0\tau)\langle\epsilon(t)\epsilon(t+\tau)\rangle\,.
 \end{equation}
 The one-sided power spectrum is defined so that
 \begin{equation}
     \int_0^\infty{\rm d}\omega S_k(\omega)
     =\langle\epsilon(t)^2 \rangle=\epsilon_0^2 \,.
 \end{equation}
Assuming that the levitated particle occupies the state $|n\rangle$ with a probability $P(n,t)$ at time $t$, the average damping rate can be calculated as
\begin{eqnarray}
    \frac{{\rm d}\langle E(t)\rangle} {{\rm d}t}
    &=&\sum_{n} P(n,t)2\hbar \omega_0 
    (R_{n\to n-2}-R_{n \to n+2})\nonumber\\
   & =& \frac{\pi}{2}\omega_0^2 S_k(\omega_0)
    \langle E(t)\rangle\,.
    \label{evolve}
\end{eqnarray}
The average energy of the oscillator, denoted as $\langle E(t)\rangle$, can be expressed as $\langle E(t)\rangle=\sum_{n} (n+\frac{1}{2})P(n,t)2\hbar \omega_0$. It is evident that the damping coefficient can be extracted from the above equation in the form,
\begin{equation}
    \frac{{\rm d}\langle E(t)\rangle} {{\rm d}t} =-\gamma_{\rm RE}\langle E(t)\rangle,
\end{equation}
where
\begin{equation}
    \gamma_{\rm RE} = \frac{\pi}{2}\omega_0^2 S_k(\omega_0)\,.    
\end{equation}

Note that the derivation of the damping coefficient in this work differs from that of Ref.~\cite{PhysRevA.58.3914}, where the heating rate is calculated. The rationale behind this work is based on the principle that every damping and fluctuation adhere to the fluctuation and dissipation theorem. By applying Eq.~\eqref{general-core}, the fluctuation correlation function, one can determine the fluctuation (or the heating rate as in Ref.~\cite{PhysRevA.58.3914}). Consequently, the damping coefficient should be calculated first, which in turn determines the fluctuation. Here, we will refer to the damping coefficient as
\begin{equation}
    \gamma_{\rm RE} = \omega_0^2 S_{\rm RE} \,.
    \label{gamresk}
\end{equation}
in which $S_{\rm RE}$ is a newly defined parameter that absorbs all the constants, including $\epsilon_0^2$ present in the expression. In the levitation experiment, we assume that $S_{\rm RE}$ can be finely adjusted to remain nearly constant within the vicinity of $\omega_0$. This allows for the straightforward extraction of the photon recoil contribution to the signature.

The recoil temperature $T_{\rm RE}$ can be derived as follows. The force fluctuations acting on the levitated particle and their spectral density, as given in Eq.~\eqref{general-core}, can be determined using the Wiener-Khinchin theorem
\begin{eqnarray}
S_{F_i F_j}(\omega)&=&\int_{-\infty}^{\infty}\left\langle\hat{F}_i(\omega) \hat{F}_j^*\left(\omega^{\prime}\right)\right\rangle d \omega^{\prime}
\label{sfifj}\\
&=&\frac{1}{2 \pi} \int_{-\infty}^{\infty}\left\langle F_i(t) F_j\left(t+t^{\prime}\right)\right\rangle \mathrm{e}^{i \omega t^{\prime}} d t^{\prime}\nonumber\\
&=&\frac{\delta_{ij}m\gamma_{\rm RE}k_{\rm B} T_{\rm RE}}{\pi}\,,
\nonumber    
\end{eqnarray}
where $\hat{F}_i(\omega)$ represents the Fourier transform of $F_i(t)$. The force exerted on the particle by a focused laser with frequency $\Omega$ is given by $F_i(\Omega)=P{\mathrm{scatt}}^{i}(\Omega) / c$, where $P_{\text {scatt }}^{i}$ represents the power scattered in the direction $i$. Therefore,
\begin{equation}
    \left\langle\hat{F}_i(\Omega) \hat{F}_j^*\left(\Omega^{\prime}\right)\right\rangle
    = \frac{1}{c^2}\left\langle\hat{P}_i(\Omega) \hat{P}_j^*\left(\Omega^{\prime}\right)\right\rangle\,.
\end{equation}
The dominant source of the fluctuation is the short noise, 
thus the power spectral density is 
~\cite{schottky1918spontane}
\begin{equation}
    S_{F_i F_j}(\Omega)=\frac{\hbar \Omega }{2\pi c^2 }
    P_{\rm scatt}^{ij}(\Omega)\,.\label{sfifj2}
\end{equation}
Compared it with the density  Eq.~\eqref{sfifj}, we can get
the effective temperature of the recoil of the quanta
$\hbar \Omega $ in $i$ direction
\begin{eqnarray}
    T_{\rm RE} &=& \frac{\hbar \Omega }{2m c^2 k_{\rm B} \gamma_{\rm RE} }
    P_{\rm scatt}(\Omega)\\
    &=& \frac{\hbar \Omega }{2m c^2 k_{\rm B} \omega_0^2 S_{\rm RE} }
    P_{\rm scatt}(\Omega)\,.    \nonumber
\end{eqnarray}
There exists a natural linear relationship between $T_{\rm RE}$ and $P_{\rm scatt}$. As discussed earlier, $T_{\rm RE}$ does not necessarily equal the temperature of the environment.

\subsection{The feedback optical force and damping rate}
In this subsection, the feedback type damping and cooling
forces in the levitation
experiments are discussed. With the continuous fast developments 
in this field,  only the essentials of the cooling 
techniques can be shown.  Feedback technology is a very important way to actively detect the speed and position of particles, communicate through circuits, and use mechanical means to give particle recovery force. Many articles have already introduced the theory of feedback mechanism and provided continuous experimental improvements~\cite{PhysRevLett.109.103603, Magrini_2021, Kamba:22, PhysRevResearch.4.033051, PhysRevA.107.023516, Guo:2023xmd, Gosling:2023lgh}.

In order to cool down the levitated particle, several detectors are used to detect the center-of-mass motion of the trapped nanoparticle from three orthogonal directions as shown in the sketch map in panel A of Fig.~\ref{sec2-lv1}. It needs to obtain a signal at twice the oscillation frequency. The detectors are responsible for obtaining the position ($z(t)$) and velocity ($\dot{z}(t)$) information of the nanoparticle. Multiply them $z(t)\dot{z}(t)$ that we can get the signal to describe the state of particle's motion and apply the required damping force or force to amplify the motion based on it. 
Considering all the feedback signals in three directions,  different devices can be used. For example, as shown in Ref.~\cite{PhysRevLett.109.103603},
the driving of a Pockels cell can modulate the power $P$ to change the transverse trap stiffness $k_{\rm{trap}}$ which has the relationship with the optical force $F_{\rm{opt}}=\Delta k_{\rm{trap}}z(t)\propto \Delta P z(t)$. 
The more precise the feedback force, the greater the corresponding equivalent damping rate, which can result in lower temperatures. There are several aspects that can affect the feedback force's ability to cool down, like the standard quantum limit (SQL) leading the measurement uncertainty of the motion of the nanoparticle. The increasing signal power can reduce the measurement uncertainty, but the scattering between the photon and the nanoparticle can also get an enhancement heating the nanoparticle. 
And what's more, the detection bandwidth affect the position accuracy,
of which the details can be found in~\cite{PhysRevLett.109.103603} and its supplement materials.

As for the theoretical formulation of the feedback mechanism, 
the damping force and optical cooling force could be used and 
these two force appear in different sides of the Eq.~\eqref{scho}.
Before going into the detail, We show the differences between the work of Ref.~\cite{2014NatNa...9..358G} and Ref.~\cite{Jain2016DirectMO}. This serves as one of the motivations for our work. Ref.~\cite{2014NatNa...9..358G} studied the levitodynamics and relaxation of a levitated particle, where cooling is achieved through the equilibrium between stochastic fluctuations and the optical damping force $F_{\rm FB}=-\omega_0\eta q^2 p$. However, as shown in Ref.~\cite{Jain2016DirectMO}, when the feedback loop is activated, an additional damping coefficient $\gamma_{\rm FB}$ appears in the Langevin system. It is conceivable that cooling can be achieved even without the presence of the optical damping force $F_{\rm FB}$. The details of the cooling are already presented in Fig.~\ref{sec2-col} and the corresponding discussion in the preceding section. Here, we provide a unified description of the two cooling effects. In fact, the feedback  effect is caused by the motion of the levitated particle, which also alters its position. If a feedback loop is set up in an experiment, neglecting the higher-order ${\cal O} (p^2)$ terms, the damping force can be generally expressed as
\begin{equation}
    F_{\rm damping} = -G(q^2 )p\,.
\end{equation}
$G(q^2)$ is a function of the square of the coordinate, $q^2$. It should be noted that the linear term in $q$ does not contribute to damping. Therefore, the leading two terms in the Taylor expansion are
\begin{equation}
    F_{\rm damping} \simeq -G(0)p-G'(0)q^2p\,.
\end{equation}
The first term represents the feedback damping coefficient, while the second term corresponds to the optical damping force. Both the feedback damping and optical damping forces exist and are deterministic. As discussed in the previous section on levitodynamics, it is important to note that the total damping $\gamma_{\rm T}$ is not necessarily equal to the stochastic damping $\gamma_{\rm S}$. However, the feedback damping can dominate the cooling process of the levitation.

Generally, every physical operation experiences some degree of disturbance
or interference from the environment. Consequently, stochastic forces should be considered in the feedback mechanism. These forces, along with the corresponding damping, can be easily accounted for by introducing a pair of terms, $\gamma_{\rm FB}^\prime$ and $F_{\rm FB}^\prime$, on both sides of Eq.~\eqref{scho}. 
As talked in above section, the theory will not be changed.
In this paper, we focus on extracting the dark damping coefficient in
the steady state without a feedback loop. Therefore, we do not present 
the detail of the parameter in feedback mechanism here.

\subsection{The collision between levitated particle and dark matter}

\begin{figure*}[htpb]
\begin{center}
\includegraphics[width=120mm]{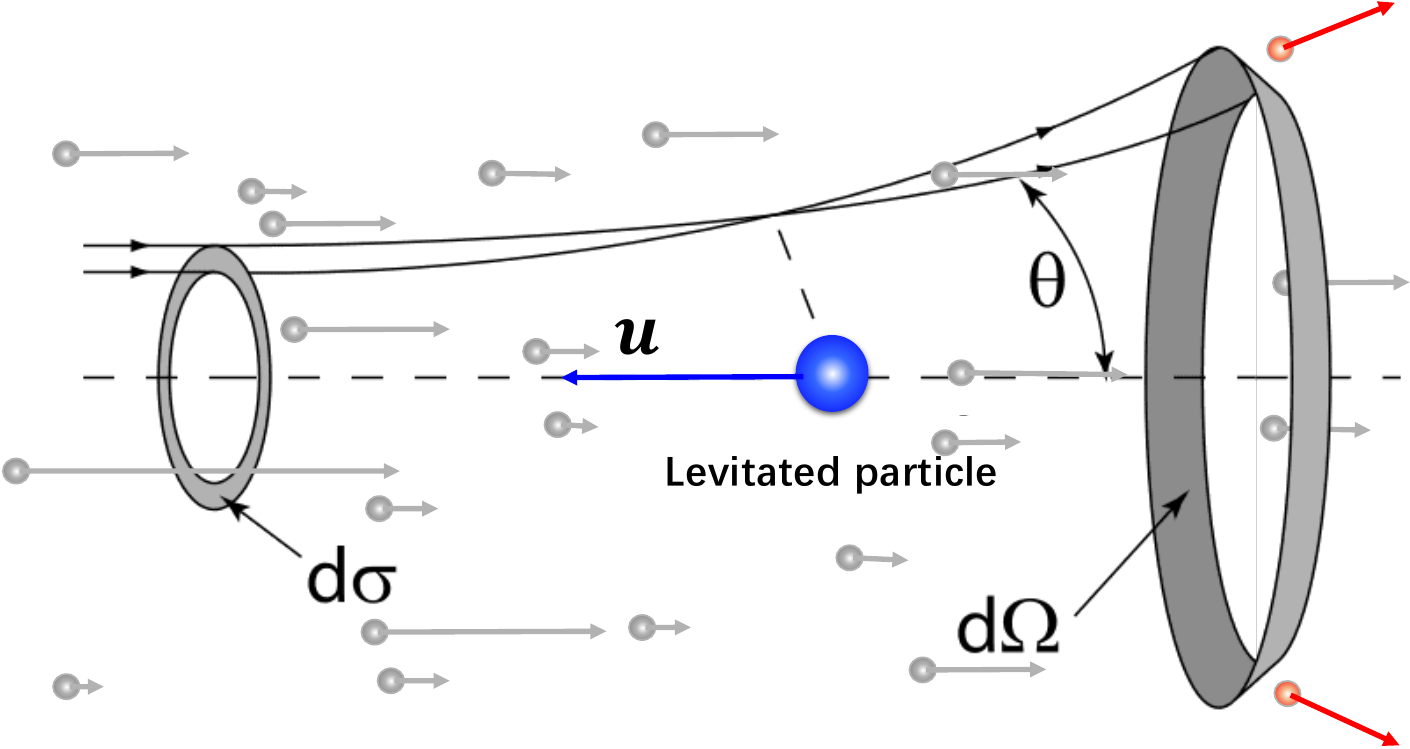}
\caption{ The sketch map of the dark matter 
scattering on a levitated particle from in one direction.
The red ball is the final scattering state of the DM. 
Counting the transfer momentum to the levitated particle gives
the damping force of the incident DM flux.}
\label{sec3-dmsc}
\end{center}
\end{figure*}

Evidence of the existence of dark matter arises from the gravitational effects observed in the behavior of galaxies and clusters, as demonstrated by various astrophysical observations. The accumulation of evidence increasingly clarifies that a significant portion of the universe's matter exists in a non-luminous form, which could be weakly interacting with elements of the standard model (SM) and thus challenging to detect in terrestrial laboratories. Direct detection aims to identify signatures of dark matter scattering off a terrestrial target in laboratory settings. Leveraging the levitation of nanoparticles provides a means to employ macroscopic force sensors for probing long-range interactions between dark and visible matter, including gravitational interactions. However, realizing such ambitious experiments would require substantial advancements beyond the current state of the art. Nevertheless, similar concepts for searching for dark matter that might interact through stronger long-range interactions are already feasible.

As talked in the introduction,
levitation sensors could possibly 
possess sufficient sensitivity to detect and
count individual collisions 
between dark matter and levitated particle. 
However, as shown in Ref.~\cite{Cheng:2019vwy},
the collision between the nanoparticle and the dark matter bath
may generate Brownian random motion too. In our work, what we
want to do is to searching the stochastic force caused by 
dark matter collision. This need to be under a 
crucial assumption that the scattering rate is sufficiently 
high, that the
time between scatters is much smaller than the inverse of the mechanical
frequency of the particle. We think current setup of levitation particle
experiments could achieve this. and this assumption will limit
the parameter space we can detection. The detailed discussion will be
give in the next section, and  this subsection gives the formulation 
for the dark damping. 
Naively, one might expect the damping rate caused by dark matter collisions to be the same as the $Kn\gg 1$ case of the isothermal drag force since the mean free path of dark matter is much greater than the size of the levitated particle. 
However, a detailed analysis reveals that damping from dark matter differs from the isothermal drag force. Although the dark matter distribution can be approximated as having a similar Maxwell-Boltzmann distribution, the linearization condition in Eq.~\eqref{eqlindis} fails as the vapor velocity of the dark matter becomes comparable to the dark matter particle velocity. More details regarding the velocity distribution can be found in Ref.~\cite{DelNobile:2021wmp, young2017survey}.

Next, we provide a brief summary of the kinetic setups of galactic dark matter particles, which are gravitationally bound to the halo of our galaxy. The local distribution of dark matter is formulated using the subhalo model of the galaxy, with a density of approximately $\rho=0.3,\rm GeV/cm^3$. Assuming only one type of dark matter particles, the numerical density of these particles near Earth decreases as the dark matter mass $m_{\chi}$ increases. The scattering rate of dark matter depends on time due to variations in the dark matter flux on Earth caused by the Earth's motion around the Sun. Consequently, the dark matter signals are expected to exhibit annual modulation. The velocity of dark matter at Earth's location is anticipated to be a few hundred km/s, limited by the galactic escape velocity. In the galactic rest frame, the velocity distribution follows the Maxwell-Boltzmann form, with the most probable velocity $v_0$ typically chosen as $220$ km/s. While the circular velocity of the Sun around the galaxy's center is approximately $240$ km/s, and the circular velocity of the Earth around the Sun is about $30$ km/s. Although the relative velocity of the levitated particle with respect to Earth can be much smaller, the drift velocity of the dark matter flux is several hundred kilometers per second. Therefore, the motion of the solar system implies that the linearization of the expansion of the drift wind $U_\infty$ in fluid dynamics, as in Eq.~\eqref{eqlindis}, is no longer applicable. Defining the pressure of the dark matter flux also becomes challenging. Consequently, we need to find an alternative approach to calculate the damping coefficient caused by dark matter.

Indeed, the picture of a saturated particle in a liquid, along with the concept of isothermal drag, belongs to the realm of macroscopic physical systems within fluid dynamics. On the other hand, when considering dark matter scattering on the levitated particle, we need to delve into concepts from fundamental particle physics, such as scattering cross sections, particle masses, and coupling strengths. These concepts are formulated within the framework of quantum scattering theory. While there are similarities between these two physical pictures, a direct calculation from quantum scattering theory to the macroscopic levitated particle is required. Therefore, we move away from fluid approaches in the subsequent study.

The damping coefficient can be straightforwardly derived from the collisions between the dark matter particle and the levitated particle. Let's consider a one-dimensional collision as an example to illustrate the details. As depicted in Fig.~\ref{sec3-dmsc}, both the levitated particle and the ambient dark matter are moving along the $x$ axis. The velocity of the dark matter particle in the flux follows a Maxwell-Boltzmann distribution.
\begin{equation}
    f(v_x,T_{\rm DM}) = \left(\frac{m_{\chi}}{2\pi k_{\rm B} T_{\rm DM}}\right)^{\frac{1}{2}}
    \exp\left(-\frac{m_{\chi}v_{x}^{2}}{2k_{\rm B} T_{\rm DM}}\right)\,.
\end{equation}
Here the  effective temperature  corresponds
to the average velocity $v_0$ of the DM 
\begin{equation}
T_{\rm DM}=\frac{\pi m_\chi}{8k_{\rm B}}v_0^2\,.\label{deftdm}
\end{equation}
The levitated particle will experience a damping force resulting from the transfer of momentum in the opposite direction from the dark matter particles. The strength of this force should be proportional to the velocity of the levitated particle, denoted as $u$, assuming it is much smaller than the velocities of the dark matter particles. However, it is important to note that in this context, $u$ cannot be equated to $U_\infty$ as in Eq.~\eqref{eqlincon} of fluid dynamics.

In Fig.~\ref{sec3-dmsc}, it is evident that the momentum transferred to the levitated particle can be computed by considering the scattering of dark matter particles within the final solid angle ${\rm d}\Omega$. The number density of the incident dark matter particles is given by $\rho_\chi/m_\chi$. Utilizing the differential elastic cross section ${\rm d}\sigma/{\rm d}\Omega$, the momentum transferred to the levitated particle during a time interval $\Delta t$ can be expressed as
\begin{equation}
\begin{aligned}
\Delta P & =\int\int\frac{\rho_{\chi}}{m_{\chi}}\left(\frac{m_{\chi}}{2\pi k_{\rm B}T_{{\rm DM}}}\right)^{\frac{1}{2}}
\exp\left(-\frac{m_{\chi}v_{x}^{2}}{2k_{\rm B}T_{{\rm DM}}}\right)\\
&\times\frac{{\rm d}\sigma}{{\rm d}\Omega}\left|v_{x}-u\right|\Delta t\times m_{\chi}v_{x}\left(1-\cos\theta\right){\rm d}v_{x}{\rm d}\Omega\\
  & =-\rho_{\chi}\Delta t\left(\frac{m_{\chi}}{2k_{\rm B}T_{{\rm DM}}}\right)^{-1}\frac{1}{4}\left(1+\text{Erf}\left(\sqrt{\frac{m_{\chi}}{2k_{\rm B}T_{{\rm DM}}}}u\right)\right.\\
  &\left.-\text{Erfc}\left(\sqrt{\frac{m_{\chi}}{2k_{\rm B}T_{{\rm DM}}}}u\right)\right)\int\frac{{\rm d}\sigma}{{\rm d}\Omega}\left(1-\cos\theta\right){\rm d}\Omega\\
 & \simeq-\rho_{\chi}\Delta t\left(\frac{m_{\chi}}{2k_{\rm B}T_{{\rm DM}}}\right)^{-1}\frac{\sqrt{\frac{m_{\chi}}{2k_{\rm B}T_{{\rm DM}}}}u}{\sqrt{\pi}}\sigma_{T}\\
 & =-\rho_{\chi}u\Delta t\left(\frac{\pi m_{\chi}}{2k_{\rm B}T_{{\rm DM}}}\right)^{-\frac{1}{2}}\sigma_{T}\,.
 \end{aligned}
 \label{deltapt}
\end{equation}
In the last line of Eq.~\eqref{deltapt}, we employ Taylor expansion to obtain the dominant term of the transferred momentum. 
It should be noted that the levitation experiments are actually conducted under the high-speed drift wind of dark matter, which results in anisotropic momentum transfer. The assumption of isotropic incident dark matter is an approximation, and the effects of anisotropy will be discussed in future work.
Here, $\sigma_T$ represents the transfer cross section defined for collisions between dark matter  particles and the levitated particle
\begin{equation}
    \sigma_T = \int\frac{{\rm d}\sigma}{{\rm d}\Omega}\left(1-\cos\theta\right){\rm d}\Omega\,.\label{tsigmat}
\end{equation}
We can see that the damping coefficient in one dimension is 
\begin{equation}
\gamma_{{\rm DM}}^{\rm 1D}=-\frac{\Delta P}{\Delta t}\frac{1}{m u}=\frac{\rho_{\chi}}{m}\left(\frac{\pi m_{\chi}}{2k_{\rm B}T_{{\rm DM}}}\right)^{-\frac{1}{2}}\sigma_T\,.\label{P1dre}
\end{equation}
The 3D result of $\gamma_{\rm DM}$ can be obtained by using the aforementioned result. The momentum transfer from the dark matter  particles to the levitated particle within an infinitesimal solid angle ${\rm d}\Omega'/2\pi$ on a hemispherical surface can be approximated using the one-dimensional result given by Eq.~\eqref{P1dre}. By integrating over ${\rm d}\Omega'/2\pi$ for all momentum components along the direction of motion of the levitated particle, we can derive the 3D result as follows
\begin{eqnarray}
&& \gamma_{\rm DM}=\int \gamma^{\rm 1D}_{\rm DM}\cos^2\theta' 
\frac{{\rm d}\Omega'}{2\pi}\label{gammadm}\\
&& =\frac{\gamma^{\rm 1D}_{\rm DM}}{3}
 =\frac{\rho_{\chi}}{3m}\left(\frac{\pi m_{\chi}}{2k_{\rm B}T_{{\rm DM}}}\right)^{-\frac{1}{2}}\sigma_T \,.    \nonumber
\end{eqnarray}
Substituted the effective temperature $T_{\rm DM}$ Eq.~\eqref{deftdm},
we can get a very simple expression of the damping coefficient
\begin{equation}
    \gamma_{\rm DM}=\frac{\rho_\chi v_0}{6m}\sigma_T\,.
    \label{gdmsigt}
\end{equation}
The damping coefficient mentioned above is suppressed by the mass of the levitated particle. However, as mentioned earlier, the transfer scattering cross section $\sigma_T$ represents the scattering between dark matter  and a macroscopic particle. The connection between this interaction and the interaction between DM and fundamental particles can be elucidated as follows. 

When comparing with the isothermal drag force discussed in the previous subsection, we observe that the transfer cross section $\sigma_T$ in Eq.~\eqref{gdmsigt} corresponds to the area $R_0^2$ in Eq.~\eqref{gamthpg}. This area provides a macroscopic description of the scattering process.
The scattering cross section and the mass of the DM are typically the two most significant parameters in dark matter detection. However, in the case of the levitation experiment, there are some misunderstandings regarding the interactions between the DM and the levitated particle due to the composite nature of the levitation system.
The levitated particle is composed of millions of microscopic molecules or atoms, depending on the mass scale (the amount of the microscopic particle) of the levitation. The detection of levitation aims to observe the collective motion of these composite particles. Therefore, the differential cross section in Eq.~\eqref{deltapt} and the transfer cross section $\sigma_T$ in Eq.~\eqref{tsigmat} do not represent fundamental dark matter interactions as commonly understood in the literature, which refer to interactions between elementary particles.

To establish the relationship between fundamental interactions and levitation interactions, we need to derive this connection. The key aspect of this derivation lies in the coherence that exists from the fundamental interaction to the levitation level. The collective movement of the levitated particles can enhance the scattering rate through coherent effects. This enhanced scattering rate gives rise to a characteristic scattering pattern known as the static structure factor
(defined below). The static structure factor results from the collective interference of waves scattered by particles in the system. This interference is sensitive to the relative separation between the particles, and the static structure factor can be expressed as the spatial Fourier transform of the particle structure, represented by the density-density correlation function.

\begin{figure*}[htpb]
\begin{center}
\includegraphics[width=150mm]{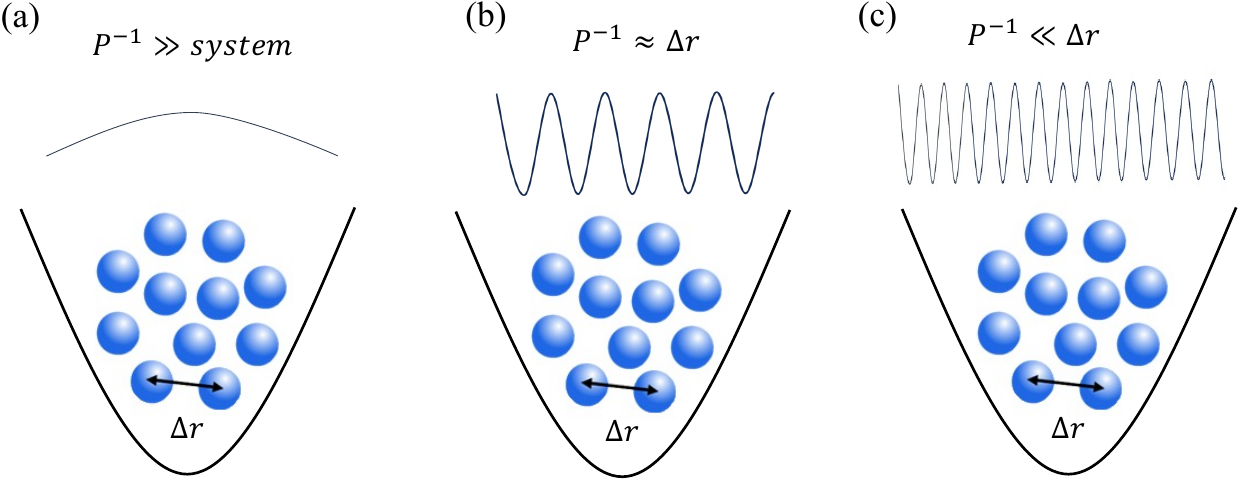}
\caption{ $S(P)$ exhibits three distinct regimes depending on the particle spacing in relation to the scattering length. In the case where the scattering length is significantly larger than the particle collection, interference manifests as a sum of nearly equivalent phases, resulting in a proportional relationship between $S(P)$ and the amount  of particles. When the scattering length is of comparable magnitude to the particle spacing, significant angular variations arise in the scattered intensity. In the scenario where the scattering length is significantly smaller than the particle spacing, the phases become randomized, resulting in interference that causes $S(P)$ to approach unity, with the $P$-dependence solely determined by the particle form factor.}
\label{sec3-sc}
\end{center}
\end{figure*}

Let's consider the dark interaction with the nucleon as an example. The scattering rate can be derived from the amplitude ${\cal M}_{\chi n}$, which originates from the fundamental dark interaction. The scattering of the levitated particle involves atoms, which are composites composed of nucleons and electrons. Thus, the total scattering is determined at two levels: the distribution of nucleons within the atom and the configuration of atoms within the levitated particle. This physical picture resembles X-Ray diffraction in a crystal, as depicted in Fig.~\ref{sec3-sc} (Ref.~\cite{sidebottom2012fundamentals}). Firstly, the spatial distribution of matter can be formulated as the nucleus form factor $f(P)$ in momentum space, which is obtained by performing the Fourier transformation of the spatial distribution.
\begin{equation}
    f(P) = \int \rho(\bm r)e^{-i\bm P \cdot \bm r} {\rm d}^3r\, .
\end{equation}
It should be noted that an isotropic matter distribution is assumed in
this case, simplifying the situation. Therefore, $f(P)$ is solely
dependent on $P$. Typically, $f(P)$ is normalized to unity to
facilitate the calculation of the total scattering amplitude 
with the atom. Also note that
this calculation involves counting the atomic number $A$.
\begin{equation}
    {\cal M}_{\chi A} = {\cal M}_{\chi n}Af(P)\,. 
\end{equation}
${\cal M}_{\chi n}$ could originate from a dark interaction between dark matter particles and nucleons, such as a scalar portal or a dark U(1) interaction. For the purposes of this initial study, it can be approximated as a constant. 

Next, we proceed to calculate the scattering on the levitated particles. In this context, the term ``levitated particle" represents a generic term encompassing the objects comprising condensed matter. Regardless, the atoms within the collection occupy distinct relative positions, giving rise to interference during the levitation process, particularly in the context of DM scattering. The displacement of the atom can be expressed as an amplitude, incorporating an additional unitary transformation $\hat U=\exp{(-i\bm r\cdot \bm P)}$, yielding the total amplitude as a result
\begin{equation}
    {\cal M}_{\chi L} = \sum_{i}^{\text{all the atoms}}
    {\cal M}_{\chi n} Af_i(P)e^{-i\bm P \cdot \bm r_i}\,.
\end{equation}
Here, ${\bm r}_i$ represents the position of the center of the $i$th nucleus. From the formulations above, we can establish a relation between the fundamental interaction and the macroscopic cross section as utilized in Eq.~\eqref{gammadm} from the damping
\begin{eqnarray}
    \sigma_T &\propto& \left|{\cal M}_{\chi n}\right|^2A^2 \\
    && \times \left(\sum_i f_i({ P}) \mathrm{e}^{-i {\bm  P} \cdot {\bm r}_i}\right)\left(\sum_j f_j({ P}) \mathrm{e}^{-i {\bm P} \cdot {\bm r}_j}\right)^\ast\,.\nonumber
\end{eqnarray}
Integrating $\left|{\cal M}_{\chi n}\right|^2$ yields the standard scattering cross section $\sigma_T^{\rm D}$ commonly studied in the literature on dark matter. Assuming all atoms are identical and possess identical form factors.
\begin{eqnarray}
\sigma_T&=&\sigma_T^{\rm D}|f({ P})|^2 A^2 N\left\{\frac{1}{N} \sum_i \sum_j \mathrm{e}^{-i {\bm P} \cdot\left({\bm r}_i-{\bm r}_j\right)}\right\}\nonumber\\
&=&\sigma_T^{\rm D}\left|f({ P})\right|^2 A^2 N S({P})\,,
\label{sigttas}    
\end{eqnarray}
where $N$ is the total number of the component particle in the levitated
particle. The static structure factor is defined by
\begin{equation}
S({ P})\equiv \frac{1}{N}\left\langle\sum_i \sum_j \mathrm{e}^{-i {\bm P} \cdot\left({\bm r}_i-{\bm r}_j\right)}\right\rangle\, ,   \label{sfact}
\end{equation}
where the angled brackets indicate an average taken over appropriate ensembles of the structure.
Note that the static structure factor $S(q)$ defined here represents an unknown interaction between nucleons and dark matter, and is therefore not the conventional static structure factor found in XRD diffraction.

The quantity $S( P)$ in Eq.~\eqref{sfact} serves as an overall measure of the phase differences in the scattered field, which arise due to the relative separation of the particles. In this context, the scattering wave vector plays a crucial role. The reciprocal of the scattering wave vector represents a significant scattering length scale, given by $l=2 \pi / P$, which determines the severity of interference effects in comparison to the mean particle spacing. This is illustrated in Fig.~\ref{sec3-sc}, which displays three distinct regimes. When the scattering length scale is significantly larger than the distances between atoms within the system (as shown in the left panel of Fig.~\ref{sec3-sc}), the phase differences, ${\bm P} \cdot\left({\bm r}_i-{\bm r}_j\right)$, between neighboring scattered waves are nearly identical, leading to constructive interference.
\begin{equation}
S({ P}) = \frac{1}{N}\left\langle\sum_{i, j} \mathrm{e}^{-i 
{\bm P} \cdot\left({\bm r}_i-{\bm r}_j\right)}\right\rangle \approx \frac{N^2}{N}=N .
\end{equation}
Conversely, when the scattering length scale is small compared to the 
particle spacing, (right panel of Fig.~\ref{sec3-sc})
the phase differences, 
${\bm P} \cdot\left({\bm r}_i-{\bm r}_j\right)$,
between waves scattered by neighboring atoms are randomized 
and produce
\begin{equation}
S({ P}) = \frac{1}{N}\left\langle\sum_{i, j} \mathrm{e}^{-i {\bm P} \cdot\left({\bm r}_i-{\bm r}_j\right)}\right\rangle \approx \frac{N}{N}=1 \, .
\end{equation}
In any case, the levitation experiments exhibit significant enhancements of the microscopic transfer cross section $\sigma_T^{\rm D}$ by factors of $A^2 N^2$ or $A^2 N$. This sensitivity to the microscopic forces is the reason behind the high sensitivity of levitation. Additionally, it should be noted that the measured momentum $p$ in the levitation corresponds to the collective motion of the levitated particles.
\begin{equation}
    P=\frac{p}{NA}\sim \frac{\sqrt{m\langle E\rangle}}{NA}\,.
\end{equation}
In this scenario, the measured energy should be significantly larger than the energy scale $\hbar \Omega$. Consequently, the nucleus form factor $f(P)$ can be safely approximated as $f(0)$, and the static structure factor can also be approximated as $N$. More details about the form factor can be found in~\cite{Riedel:2012ur}. Further studies on DM detection are discussed in the following section.

Substituted Eq.~\eqref{sigttas} to 
Eq.~\eqref{gdmsigt}, the dark damping coefficient is
\begin{eqnarray}
    \gamma_{\rm DM}&=&\frac{\rho_\chi v_0}{6m}\sigma_T^{\rm D}A^2 N|f({P})|^2 S({P})\nonumber \\
    &\approx &\frac{\rho_\chi v_0}{6m_{\rm proton}}\sigma_T^{\rm D}A |f({P})|^2 S({P})\,.
    \label{gdmehn}
\end{eqnarray}
Note that we simply assume that $m\approx m_{\rm proton}AN$ in above
equation. It is evident that the damping coefficient, despite being suppressed by the mass of the levitated particle, is enhanced by the amount of fundamental particles. When both $f(P)\to 1$ and $S(P)\to N$ occur, the coefficient is enhanced to the point of resembling a macroscopic interaction. This distinction arises from the coherence of the scattering and sets apart dark damping from all other types of damping discussed in the preceding subsections.

Comparing all the discussed damping coefficients in this section, we can observe that the distinct properties of these coefficients provide valuable clues for extracting each damping effect and its corresponding fluctuation from specific experimental setups and appropriate conduction in levitation experiments. In the next section, we revisit the initial motivation behind our work, which is to extract the dark interaction from the aforementioned damping effects. We aim to provide an estimation of the DM parameters, including the cross-section and interaction strength, among others.

\section{Exploration on the dark damping in the levitation experiments}
\label{sec4-dm}

\subsection{The linear response and the extraction 
of the damping coefficient}

Before delving into the experimental exploration, it is crucial to
ascertain the reactions that occur in the laboratory. Typically, 
the linear response of the levitated particle to driving forces 
is assumed. Subsequently, the response function can be derived 
as a perturbation of the system. The analyticity, causality, 
and the Kramers-Kronig relation in the response can be found 
in standard textbooks on general kinetic theory~\cite{pathria2016statistical}.
It should be noted here that with the inclusion of the feedback force $F_{\rm FB} = -\omega_0 \eta q^2 p$, the levitated system is no longer a linear response system. However, in the investigation of dark damping, the feedback loop is turned off, and we propose to measure the linear response in the equilibrium state. Therefore, the linear response relations remain applicable. Here we give the essentials for
the reaction and dissipation based on Eq.~\eqref{laneq},
incorporating a general damping coefficient ($\gamma$) and 
temperature ($T$). 

For the case of natural fluctuations in the position and velocity 
of the particle in equilibrium, the stochastic force averages to 
zero and is assumed to have delta-function time correlation, 
as depicted in Eq.~\eqref{general-core}. The average of 
the position and velocity squares obeys the equipartition theorem
\begin{equation}
    \langle q^2 (t)\rangle=\frac{k_{\rm B} T}{m\omega_0^2},~~
    \langle p^2(t)\rangle=m k_{\rm B} T\,.
\end{equation}
The power spectrum of $q^2$ is the Lorentzian peak 
\begin{equation}
    \tilde S_{q^2}(\omega)=\frac{2\gamma k_{\rm B} T}{m}
    \frac{1}{(\omega_0^2 -\omega^2)^2+\gamma^2 \omega^2}\,.
\end{equation}
This spectrum can be measured by counting the occupation number in a 
levitation experiment. The dissipative part of the response function
is related to
\begin{equation}
    {\rm Im}\tilde\chi (\omega)=\frac{m}{2k_{\rm B} T}\tilde S_{q^2}(\omega)\,.
\end{equation}
This result suggests that the dissipation caused by driving a system out of equilibrium with an external force is proportional to the power spectrum of the natural fluctuations that arise in equilibrium.

\begin{figure}[htpb]
\begin{center}
\includegraphics[width=65mm]{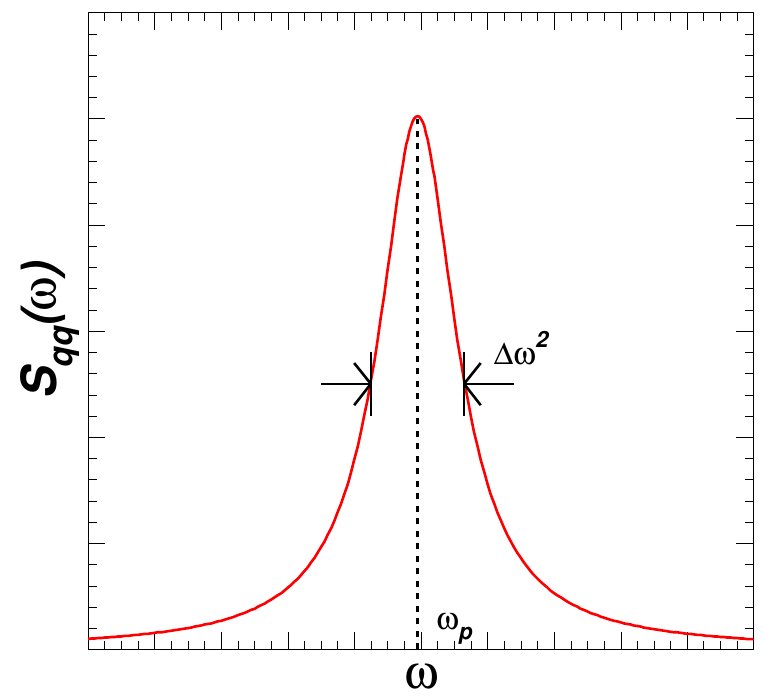}
\caption{The power spectrum $\tilde S_{q^2}(\omega)$ is characterized by the peak frequency $\omega_p$ and the half width $\Delta \omega^2$.}\label{sec4-gam}
\end{center}
\end{figure}

The power spectrum $\tilde S_{q^2 }(\omega )$ depicted in Fig.~\ref{sec4-gam} can be measured in levitation experiments by counting the occupation number of quanta $\hbar \omega_0$ in $\omega$-space. It provides comprehensive information that allows for the precise extraction of the damping coefficient $\gamma$ and temperature of the levitated particle. Various approaches exist for this purpose, and we adopt the simplest one. The primary measurable parameter is the peak frequency $\omega_p$, from which one can readily derive the relation.
\begin{equation}
    \omega_p^2 =\omega_0^2 -\frac{\gamma^2}{2}\,.
\end{equation}

However, the damping coefficient cannot be obtained directly from this relation since $\omega_0$ is not a precisely predicted or measured variable in real experimental studies. Therefore, in this paper, we propose measuring the half-width $\Delta \omega^2$ of the peak, which represents the difference between the squared frequencies at half the peak height. This allows us to establish the relation.
\begin{equation}
    \omega_{0}^{2}=\omega_{p}^{2}\sqrt{1+\left(\frac{\Delta\omega^{2}}{2\omega_{p}^{2}}\right)^{2}}\,,
\end{equation}
to determine $\omega_0$ more precisely. This relationship reveals that, in the presence of a very sharp peak, we can obtain an approximate damping coefficient.
\begin{equation}
    \gamma=\frac{\Delta\omega^{2}}{2\omega_{p}}\,.
\end{equation}
With the precisely determined coefficient, the temperature
can be derived at the peak
\begin{equation}
    T=\tilde S_{q^2}(\omega_p)\left(\omega_0^4 -\omega_p^4\right)
    \frac{\gamma m}{k_{\rm B}}\,.\label{sqtot}
\end{equation}
By utilizing this measured damping coefficient $\gamma$ and temperature $T$, we can conduct a comprehensive study of the correlation function and the fluctuation dissipation theorem. Additionally, we can search for evidence of dark interaction through the investigation of dark damping.

\subsection{The verification of the multi-stochastic force theory and the 
searching for the dark damping}

In this subsection, we present the experimental exploration of the theory proposed in this work and the search for dark interactions hidden in the noise of the levitation. Firstly, we list all the new ideas proposed in the above context:
\begin{enumerate}
    \item Multiple sources of stochastic force exist, each independently adhering the fluctuation-dissipation theorem. These sources collectively interact with the levitated particle.
    \item The cooling feedback mechanism provides a damping coefficient and an optical force that also damps the levitated particle. Consequently, the forces within the feedback mechanism are deterministic.
    \item The levitated particle is assigned an effective temperature. When the feedback is turned off, the effective temperature differs from the temperature of the environment and is denoted as $T_{\rm S}$, which may include contributions from dark interactions.
    \item The motion of the levitated particle arises from the collective motion of microscopic particles, thereby enabling the dark damping to unveil interactions between elementary and macroscopic particles.
\end{enumerate}
The first two items provide physical insights into levitation physics, while the other two items represent physical relations and quantities that can be experimentally verified or measured. Fortunately, most of the relations discussed in the previous section are linear, allowing the use of linear estimation methods in experimental studies. 

To demonstrate the effectiveness of linear estimation, the inequality between the stochastic temperature $T_{\rm S}$ and the environment temperature $T_{\rm E}$ is initially verified.
In the actual levitation experiment, we can disable the feedback at various environment temperatures $T_{{\rm E}_i}$ while keeping all other levitation setups, such as laser intensity, unchanged. Subsequently, we can measure $N$ samples of the stochastic temperature $T_{{\rm S}_i}$ by analyzing the power spectrum and simulating the relation given by Eq.~\eqref{sqtot}. By utilizing Eq.~\eqref{defTs}, we can define a linear relationship
\begin{equation}
    T_{{\rm S}}=\theta_1 + \theta_2 T_{{\rm E}}\,.
\end{equation}
If only $T_{{\rm E}_i}$ varies in the experiment, we can treat $\theta_1$ and $\theta_2$ as constants. Non-zero $\theta_1$ and the inequality $\theta_2\ne 1$ serve as verification for our proposal.  $\chi^2_{\rm T}$ is defined as
\begin{equation}
    \chi^2_{\rm T} =\sum_i^N\left(T_{{\rm S}_i}
    -\theta_1 - \theta_2 T_{{\rm E}_i}\right)^2\,.
\end{equation}
The Least Square Method gives the best matching result
\begin{eqnarray}
    &&\hspace{-4mm}\theta_1=
    \frac{\left(\sum_i^N T_{{\rm E}_i}^2\right)\left(\sum_i^NT_{{\rm S}_i}\right)-
    \left(\sum_i^N T_{{\rm E}_i}T_{{\rm S}_i}\right)\left(\sum_i^N T_{{\rm E}_i}\right)}
    {N\left(\sum_i^N T_{{\rm E}_i}^2\right)-\left(\sum_i^N T_{{\rm E}_i}\right)^2}\,,\nonumber \\
    &&\hspace{-4mm}\theta_2=
    \frac{N\left(\sum_i^N T_{{\rm E}_i}T_{{\rm S}_i}\right)-
    \left(\sum_i^N T_{{\rm E}_i}\right)\left(\sum_i^N T_{{\rm S}_i}\right)}
    {N\left(\sum_i^N T_{{\rm E}_i}^2\right)-\left(\sum_i^N T_{{\rm E}_i}\right)^2}\,.\nonumber
\end{eqnarray}
It is evident that $\theta_1=0$ and $\theta_2=1$ when every environment temperature is equal to the stochastic temperature $T_{{\rm E}_i}=T_{{\rm S}_i}$. Therefore, the agreement of $\theta_1$ and $\theta_2$ provides a direct test of the proposal in this work.

\begin{figure}[htpb]
\begin{center}
\includegraphics[width=75mm]{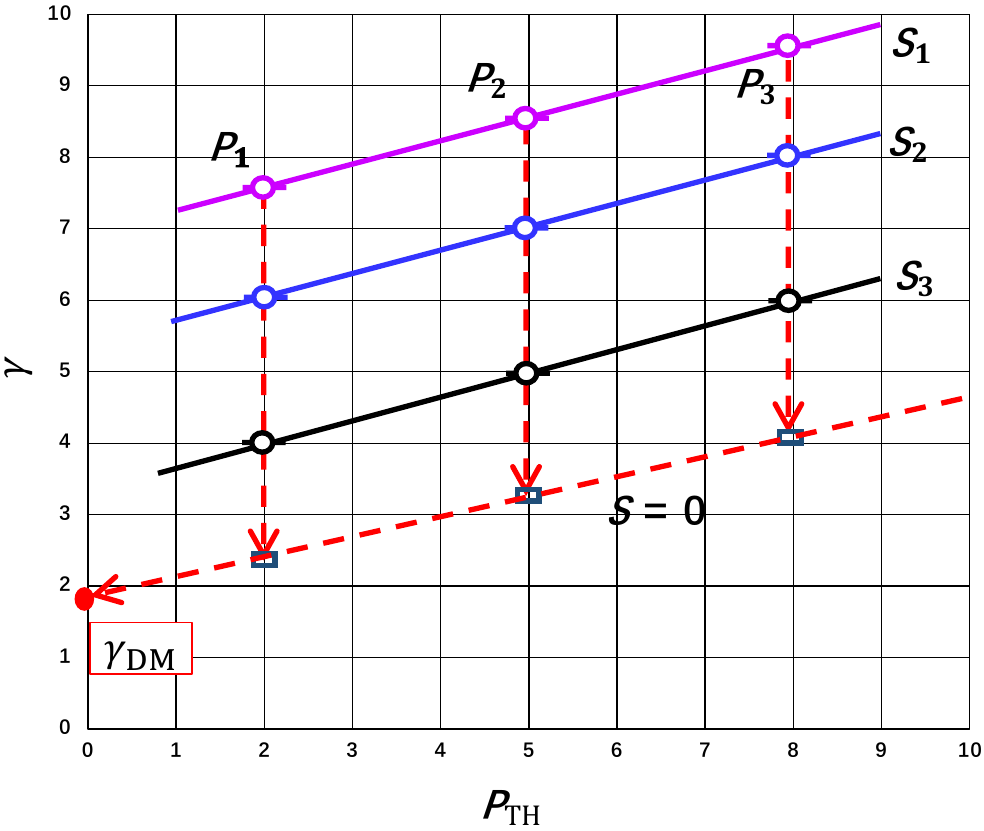}
\caption{ The sketch map using linear estimation to extract
the dark damping $\gamma_{\rm DM}$. }
\label{sec4-gamdm}
\end{center}
\end{figure}

Our objective is to search for the dark interaction in the damping coefficient. However, detecting the dark interaction requires precise tuning of the experimental setup and conduct. We propose two procedures for conducting the experiment and exploring dark damping, considering all the analyzed damping coefficients in the previous section.
\begin{enumerate}
    \item Firstly, we propose measuring the steady state without feedback in a high vacuum environment. Although cooling and heating measurements are possible, measuring the steady relaxation state should be easier and more precise. In a high vacuum environment, where the Knudsen number $Kn$ is much greater than 1, the damping coefficient $\gamma_{\rm TH}$ in Eq.~\eqref{gamthpg} is proportional to the pressure of the ambient gas. The zero pressure limit can be approached using linear estimation. Additionally, by varying the noise spectral density of the optical trapping potential, denoted as $S_{\rm RE}$ in Eq.~\eqref{gamresk}, it is possible to control the range of $\gamma_{\rm RE}$. Then, the damping coefficient and temperature can be measured, and linear estimation can be used to approach the zero optical fluctuation limit. 
    \item In the absence of optical fluctuation and pressure, non-zero damping coefficients provide upper limits on the transfer cross section between dark matter and the levitated particle. Subsequently, by utilizing the static structure factor $S(P)$ and nucleus factor $f(P)$, we can obtain constraints on the microscopic transfer cross-section between dark matter  and fundamental particles.
\end{enumerate}

Fig.~\ref{sec4-gamdm} illustrates a schematic diagram showcasing the use of linear estimation to extract the dark damping $\gamma_{\rm DM}$ from the precisely measured $\gamma_{\rm S}$. The actual experiment can be conducted under different vacuum pressures, allowing for the measurement of $\gamma_{\rm S}$ at various $S_{\rm RE}$ values. Linear estimation can then be employed to obtain the value of $\gamma_{\rm S}$ in the limit $S_{\rm RE}\to 0$. By performing linear estimation of $\lim_{{ P_{\rm TH}\to 0 }} \gamma_{\rm S}^{ S_{\rm RE}\to 0 }$, the value of $\gamma_{\rm S}$ at zero pressure can be determined. Alternatively, if $S_{\rm RE}$ can be adjusted to the same value, the linear estimation can be initially performed on the vacuum pressure, as depicted in Fig.~\ref{sec4-gamdm}. These two different procedures can serve as a cross-check to validate the final results of the damping coefficient.

The final step involves extracting constraints on the dark matter  side, such as the transfer cross section $\sigma^{\rm  D}_T$, based on the previously derived dark damping coefficient $\gamma_{\rm DM}$ from Eq.~\eqref{gdmehn}. The levitation experiment offers a high-precision sensor that enables the measurement of the coupling between a fundamental particle and a macroscopic particle. Since the focus of this paper is on studying levitodynamics and determining the dark damping based on sensor measurements, the detailed constraints on the cross-section and dark matter mass are beyond the scope of this work. For simplicity, we assume that $\sigma^{\rm  D}_T$ is constant, the nucleus form factor $f(P)=1$, and the static structure factor $S(P)=N$. Additionally, we assume identical couplings between the dark matter and nucleons. By substituting the DM density, solar system velocity, and mass of the levitated particle into Eq.~\eqref{gdmehn}, the transfer cross section can be obtained.
\begin{equation}
    \sigma_{T}^{\rm D}  \lsim  
    \frac{6\gamma_{\rm DM}}{220\times 10^5 {\rm s}^{-1}}\frac{m_{\rm proton}^2 } {0.3{\rm GeV}\times m_{\rm levitated~particle}}~
    {\rm cm^2 }\,.\label{eq:limsig}
\end{equation}
A few comments on the our exploration of the dark damping should
be addressed
\begin{enumerate}
    \item The detection limits. As talked in the above section, The 
    detection of dark damping should be under an assumption that the scattering rate is sufficiently high, that the time between scatters is much smaller than the inverse of the mechanical frequency of the particle. While the scattering rate depends on the density of the
    dark matter and the cross sections. One can get the similar
    requirement following the derivation of the Eq.~\eqref{gdmehn} 
    \begin{eqnarray}
        \frac{1}{\frac{\rho_\chi}{m_\chi}v_0 \sigma_T^{\rm D}A^2 N|f({P})|^2 S({P})}
        \ll \frac{2\pi }{\omega_0}\,.        
    \end{eqnarray}
    The denominator of the left side of the upper equation is 
    the estimated scattering rate. This in fact set a limit on
    the detection which is 
    \begin{eqnarray}
    \frac{\sigma_T^{\rm D}}{m_\chi }\gg\frac{\omega_0}
    {2\pi \rho_\chi v_0 A^2 N|f({P})|^2 S({P})}\,.
    \end{eqnarray}
    To give a constant damping effect, the scattering rate also should
    be much greater that the dark damping rate. Compared with
    the $\gamma_{\rm DM}$ in Eq.~\eqref{gdmehn}, we can see that
    it requires  $m_\chi\ll m_{\rm proton}$. This meets the initial
    motivation talked in the introduction that Levitation experiment may give us a perfect strategy to search for the sub-GeV
    dark matter.
    \item The key point of the this work is to separate a possible stochastic force from dark matter from all other sources of unknown technical noise in highly precision measurement of levitated particle experiment. Annual or daily modulation is the typical feature of the dark matter signal which can distinguish
    the dark damping from other noises. We see from the above
    section the damping coefficient is in fact the  
    factor proportional the relative velocity of the levitated
    particle in the dark matter drift wind. Thus the modulation 
    of effect can appear in nucleus form factor $f(P)$ and 
    the static structure factor $S(P)$ due to the annual 
    or daily modulation on the  transfer momentum $P$. Cross
    check can be done by comparing these factor in X-ray diffraction
    in a crystal.
    \item As talked above, the incident dark matter are
    assumed as isotropic distribution. However, the
    particle are actually levitated under the high-speed
    drift wind of dark matter. Therefore in the actual levitation
    experiment, the kinetic energy transfer from the 
    the drift dark matter wind can give a constant force acting
    on the levitated particle. To be more clear, 
    along the direction of the drift wind, one can get the force
    \begin{eqnarray}
       F= && \rho_{\chi}\left(\frac{m_{\chi}}{2\pi k_{\rm B}T_{{\rm DM}}}\right)^{-\frac{1}{2}}\sigma_{T}\\
       && \times    
       \left[\int_{-u}^{u}\exp\left(-\frac{m_{\chi}v_{x}^{2}}{2k_{\rm B}T_{{\rm DM}}}\right)\upsilon_{x}^{2}{\rm d}\upsilon_{x}\right.\nonumber\\
       && ~~\left.+2u\int_{u}^{\infty}\exp\left(-\frac{m_{\chi}v_{x}^{2}}{2k_{\rm B}T_{{\rm DM}}}\right) v_{x}{\rm d}\upsilon_{x}\right]\,.\nonumber
    \end{eqnarray}
    Here  $u$ is the large velocity of levitated particle 
    in the drift dark matter wind. 
    The force $F$ will be balanced by levitation setup such as the 
    optical forces, thus it shift the oscillation to 
    a new equilibrium point. The damping coefficient
    along this direction can be extract from small variation of
    $\Delta u$. while the coefficient along the direction 
    perpendicular to the drift wind will be the same as 
    Eq.~\eqref{P1dre}. In such cases, 
    naively thinking, the $\gamma_{\rm DM }$ is averaged over the duration of the measurements. Thus if the measurements on the power spectrum of the noise can be done in daily time or hour time, the modulation of the dark matter drift wind could give anisotropic results. Nevertheless,
    if we can measure the damping rates in different direction 
    at the same time, the anisotropic results could be double checked
    by the results measured at different time. For example, if 
    the daily anisotropy is measured, the damping coefficients
    of two orthogonal direction would be interchanged with each
    other between every six hours measurements.
\end{enumerate}
The last two comments need further deep studies beyond this work. Note
that there are other disturbations such as the scattering of the dark matter in the atmosphere or surrounding material which would cause the incident dark matter velocity distribution to differ from the assumed  distribution. These disturbations are also beyond this work and 
need our further exploration.

Though our proposal has not been applied in the experiments, the
error of current measured results in levitation experiments could
give us an estimation on the order of the non-zero value of dark
damping $\gamma_{\rm DM}$. Thus
we examine the constraints on the transfer cross-section from current levitation experiments by utilizing the results of two experiments. 
Fused silica ($\mathrm{SiO_{2}}$) particles are consistently employed in these experiments. In the experiment described in Ref.~\cite{Jain2016DirectMO}, the levitated $\mathrm{SiO_{2}}$ particles have radii on the order of 50 nm, enabling direct measurement of the photon recoil. The rate $\Gamma_{\rm RE}=\gamma_{\rm RE}n_\infty$ represents the product of the damping coefficient and the total number of quanta in the final steady state. The error in $\Gamma_{\rm RE}$ is approximately ${\cal O}(0.1)$ kHz, while $n_\infty$ is on the order of ${\cal O}(10^{5})$. Therefore, an estimation of the limit on $\gamma_{\rm DM}$ can be made, placing it on the order of ${\cal O}(10^{-3})$ Hz. 
$\gamma_{\rm DM}$ in Eq.~(87) can be replaced by this experimental error to provide a rough estimation of the limit on the transfer cross-section. Then
by inputting the corresponding parameters for fused silica, 
we can derive the limit $\sigma^{\rm D}_{ T}\lsim {\cal O}(10^{-20})\rm cm^2$. In the experiments described in Ref.~\cite{Monteiro:2020wcb}, the particle radii are on the order of $5\mu\rm m$, and the error of the Voigt profile is also on the order of ${\cal O}(10^{-3})$ Hz. Consequently, the limit on the transfer cross section could be $\sigma^{\rm D}_{ T}\lsim {\cal O}(10^{-26})\rm cm^2$. 
Note that detecting dark matter via levitation is similar to dark dynamic friction, which results from the same type of collisions between dark matter and atoms within celestial bodies and galactic gases. The limits on the DM-nucleon cross-section derived from astronomical observations~\cite{Mahdawi:2018euy,Boddy:2018wzy,Ooba:2019erm, Driskell:2022pax,Li:2022mdj,lin2023darkmattersubhaloevaporation} provide a complementary check on dark matter detection. For example, Ref.~\cite{Mahdawi:2018euy} established a limit on the p-wave transfer cross-section, corresponds to $\sigma_{T}^{\rm D} \sim 10^{-27}\ \rm{cm^{2}}$, for dark matter mass above $\mathcal{O}(1)\ \rm{GeV}$. As shown in Eq.~(87), the limit derived from levitation is independent of mass, making this detection method more practical for sub-GeV dark matter.
However, the estimation given here is very 
rough and not a actual linear estimation. Conducting an actual 
experiment to search for dark damping would yield a much more precise
limit. The linear estimation of the damping coefficient would extend
to a further non-zero damping limit, potentially serving as
evidence of dark matter. 
Another important point to mention is that neutrino 
background can also produce similar dark damping effects,
giving a similar neutrino floor as that in the WIMP detection.
For solar neutrinos, as shown in Ref.~\cite{BOREXINO:2018ohr}, the 
$pp$ solar neutrino flux is approximately $10^{11}$
$\rm{cm^{-2}s^{-1}}$, and the cross-section for neutrino 
interactions with electrons is $\sigma \sim 10^{-44}$ $\rm{cm^{2}}$.
The total inclusive charged current cross-section for 
neutrinos is $\sim 10^{-38}$ $\rm{cm^{2}}$~\cite{PhysRevD.110.030001}.
The measured damping would probably
show the existence of dark matter in the space above these limits.
More precise measurements will help us to distinguish 
the dark matter and neutrino.

\section{conclusion}\label{sec5-cl}

If the terrestrial environment is filled with dark matter, the levitation experiences contributions from damping forces and fluctuations. In this paper, we study levitodynamics with multiple stochastic forces, including thermal collision, photon recoil, feedback, and others. We assume that all stochastic forces are independent of each other, following the fluctuation dissipation theorem. We observe that the fluctuation term does not necessarily correspond one-to-one with the damping coefficient when the feedback loop is active. Therefore, we introduce two coefficients, $\gamma_{\rm T}$ and $\gamma_{\rm S}$ to differentiate between total damping and stochastic damping. The ratio of these coefficients distinguishes our levitodynamics from those with only single stochastic forces. Both feedback damping and optical damping can cool down the levitated particle. We analyze the energy distribution, effective temperature, and heating and cooling behaviors. The cooling limit differs from the result presented in Ref.~\cite{2014NatNa...9..358G}, where the limit is determined by the balance between single fluctuation and feedback optical force. In our theory, feedback damping can dominate the cooling limit. Importantly, even in an equilibrium state without feedback, the effective temperature of the system is not equal to the ambient or thermal temperature. It is a complex value influenced by thermal collision, photon recoil, and possibly dark matter.

We investigate the stochastic sources of forces in our study, including thermal drag, recoil, and feedback. These coefficients are derived from various theories such as fluid dynamics and quantum optics. We discover that DM collisions cannot be treated in the same way as in fluid dynamics due to the failure of linearizing the velocity distribution. It becomes necessary to extract the fundamental interaction between DM and standard model particles. The paper provides a detailed analysis of the relationship between the fundamental transfer cross section and the macroscopic transfer cross section. Although the dark damping coefficient is suppressed by the mass of the levitated particle, scattering can be coherently enhanced based on the amount of the component microscopic particle, the nucleus form factor, and the static structure factor. Therefore, dark damping may offer insights into detecting the macroscopic strength of fundamental particles.

Finally, we propose the operations of the levitation experiment. By precisely measuring the Lorentzian peak, a much more accurate damping coefficient can be obtained. As the thermal drag and recoil can be linearly estimated to approach zero limits, non-zero results at zero pressure and optical fluctuation can test our theory. Based on the current levitation results, we provide a rough estimate that the fundamental transfer cross section is on the order of $\sigma^{\rm D}_{T} \lesssim {\cal O}(10^{-26})\ \rm cm^2$, which is achievable only within a specific dark matter mass range. However, this is a preliminary estimate and further refinement with additional data is needed. If our proposal is implemented in actual experiments, we believe that it can provide deeper constraints and reveal more details about the dark interaction through levitation experiments.

Although this paper only focuses on four types of stochastic forces, our theory can be readily extended to scenarios involving additional stochastic forces. The linear estimation can also be expanded to encompass other relationships found in the additional damping coefficients. Extracting the dark damping from the damping coefficient is akin to removing noise in a radio telescope in the 1960s. The cosmic microwave background (CMB) was eventually discovered in the residual noise. Similarly, dark damping may serve as an irremovable noise in levitodynamics, potentially leading to the revelation of hidden secrets in the future.

\section*{Acknowledgments}
Thanks for very useful discussion with Meng Sun and Bo-Yang Liu.
This work was supported by the Natural Science Foundation of China
(NSFC) under grant numbers 12475104 and 12275232.   
\bibliography{refs}
\end{document}